\documentclass[USenglish,oneside,twocolumn]{article}
\usepackage{etex}
\usepackage[utf8]{inputenc}
\usepackage[big]{dgruyter_NEW}
\usepackage{mdwmath}
\usepackage{mdwtab}
\usepackage{eqparbox}
\usepackage{array}
\usepackage{xspace}
\usepackage[skip=1pt,font=footnotesize]{caption}
\usepackage[usenames, dvipsnames]{color}
\usepackage{float}
\usepackage{tikz} \newcommand*\circled[1]{\tikz[baseline=(char.base)]{ \node[shape=circle,draw,inner sep=1pt] (char) {#1};}}
\newcolumntype{C}[1]{>{\centering\let\newline\\\arraybackslash\hspace{0pt}}m{#1}}
\newcolumntype{L}[1]{>{\let\newline\\\arraybackslash\hspace{0pt}}m{#1}}
\expandafter\def\expandafter\normalsize\expandafter{%
	\normalsize
	\setlength\abovedisplayskip{1pt}
	\setlength\belowdisplayskip{1pt}
	\setlength\abovedisplayshortskip{1pt}
	\setlength\belowdisplayshortskip{1pt}
}
\newcommand{\privyseal}{{{\href{https://privyseal.epfl.ch}{\mbox{\tt{PrivySeal}}}}}\xspace}

\newcommand{\minisec}[1]{\subsubsection{#1:}}
\newcommand{\tparagraph}[1]{\noindent\textbf{#1:}}

\DOI{foobar}

\cclogo{\includegraphics{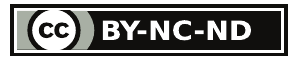}}

\begin{document}
	%
	\setlength{\textfloatsep}{10pt plus 1.0pt minus 2.0pt}
	\newcommand{\specialcell}[2][c]{ \begin{tabular}[#1]{@{}c@{}}#2\end{tabular}}

	\author*[1]{Hamza Harkous}
	
	\author[2]{Rameez Rahman}
	
	\author[3]{Bojan Karlas}
	
	\author[4]{Karl Aberer}
	
	\affil[1]{EPFL, \mbox{E-mail: hamza.harkous@epfl.ch}}
	
	\affil[2]{EPFL, E-mail: rrameez@gmail.com}
	
	\affil[3]{EPFL, E-mail: 
		bojan.karlas@epfl.ch}
	
	\affil[4]{EPFL, E-mail: karl.aberer@epfl.ch}

	\title{\huge The Curious Case of the PDF Converter that Likes Mozart: Dissecting and Mitigating the Privacy Risk of Personal Cloud Apps}
	
	\runningtitle{Article title}
	

	\begin{abstract}
		{Third party apps that work on top of personal cloud services, such as Google Drive and Dropbox, require access to the user's data in order to provide some functionality. Through detailed analysis of a hundred popular Google Drive apps from Google's Chrome store, we discover that the existing permission model is quite often misused: around \textit{two-thirds} of analyzed apps are \emph{over-privileged}, i.e., they access more data than is needed for them to function. In this work, we analyze three different permission models that aim to discourage users from installing over-privileged apps. In experiments with 210 real users, we discover that the most successful permission model is our novel ensemble method that we call \emph{Far-reaching Insights}. Far-reaching Insights inform the users about the data-driven insights that apps can make about them (e.g., their topics of interest, collaboration and activity patterns etc.) Thus, they seek to bridge the gap between what third parties can actually know about users and users' perception of their privacy leakage. The efficacy of Far-reaching Insights in bridging this gap is demonstrated by our results, as Far-reaching Insights prove to be, on average, \textit{twice} as effective as the current model in discouraging users from installing over-privileged apps. In an effort to promote general privacy awareness, we deployed \privyseal, a publicly available privacy-focused app store that uses Far-reaching Insights. Based on the knowledge extracted from data of the store's users (over 115 gigabytes of Google Drive data from 1440 users with 662 installed apps), we also delineate the ecosystem for 3rd party cloud apps from the standpoint of developers and cloud providers. Finally, we present several general recommendations that can guide other future works in the area of privacy for the cloud. To the best of our knowledge, ours is the first work that tackles the privacy risk posed by 3rd party apps on cloud platforms in such depth.}
	\end{abstract}
	
	\journalname{Proceedings on Privacy Enhancing Technologies}
	\DOI{10.1515/popets-2016-0032}
	\startpage{123}
	\received{	\vspace{-1\baselineskip} 2016-02-29} 
	\revised{2016-06-02}
	\accepted{2016-06-02}
	
	\journalyear{}
	\journalvolume{2016}
	\journalissue{4}

	\maketitle

	\section{Introduction}

	Cloud services such as Google Drive, Dropbox, OneDrive etc., have become increasingly popular in recent years. At the same time, such services have raised privacy concerns about users' data. But the danger is graver than it appears. While such cloud services are few in number, and, at least, have clearly defined privacy policies, they also serve as platforms that allow a myriad of 3rd party apps to work on top of users' data. These 3rd party apps provide certain functionalities for which they require access to the users' data. Put simply, users sacrifice some of their privacy in order to get functionalities that such apps provide. However, it appears that often such apps acquire more data than is needed for them to function.
	
	As a case study, we did an analysis of a 100 third-party apps on one of the most popular personal cloud services, namely Google Drive (240 million active users in 2014~\cite{GdriveUsers}), and discovered that almost two-thirds (64\%) of these apps require more permissions than they actually need for functioning. 
	Thus, users often end up exposing more data than is needed to unaccountable apps. For instance, a user's favorite PDF converter is highly likely to get access to her music library and discover her taste in Mozart or obtain her geo-tagged photos and know where she went on the weekend. Throughout this paper, we refer to such apps as \textit{over-privileged apps}. 
	As observed in other 3rd party apps ecosystems, giving such over-privileged apps superfluous access can potentially result in users' data being abused. This has recently been the case in the health apps market where the top 20 most visited apps were found to be sharing users' data with 70 analytics and advertising companies~\cite{healthApps}.
	
	Nevertheless, the cloud apps ecosystem has unique features that warrant a particular study of this ecosystem. Unlike studies on mobile app ecosystems, where the permissions concern the user's list of contacts, current location, or photos, the cloud permissions allow the 3rd party apps to get access to any file the user has stored in the cloud. Thus, instead of profiling the current user context, such apps can get far-reaching insights inferred from her documents, concerning her financial, legal, or health-related outlook for example. Put simply, the scale and the quality of data that can be collected is both a privacy nightmare for unaware users and a goldmine for advertisers. Second, collecting the permissions of cloud apps is challenging. Unlike other ecosystems where app permissions of thousands of apps can be easily mined via traditional web crawling, each 3rd party cloud app has a unique interface that links to the service providers. Hence, this limits the corpus of apps that one can study.
	
	In this paper, we aim to dissect and mitigate the privacy risk posed by such over-privileged 3rd party cloud apps. We study Google Drive's app ecosystem from the standpoint of all relevant parties, namely, the users, app developers and cloud providers. Generally, we seek to characterize (a)~influencing factors that can deter users from installing over-privileged apps, (b)~conditions determining developers misbehavior, and (c)~the steps cloud providers can take to mitigate users' privacy risks. To the best of our knowledge, ours is the first work that studies the privacy risk of 3rd party apps in personal cloud ecosystems. 
	
	Towards that end, we present three different permission models namely: \textit{(a)~Delta Permissions}, \textit{(b)~Immediate Insights}, and \textit{(c)~Far-reaching Insights}. The first model, i.e., \textit{Delta Permissions}, informs users about the unneeded permissions that over-privileged apps are using. 
	The second model, \textit{Immediate Insights} presents randomly selected examples from the user's data such as portions of text or image files, photo locations etc., that over-privileged apps can get access to. 
	The third model, \emph{Far-reaching Insights}, has been motivated by the novel concept of \emph{Inverse Privacy}~\cite{inversePrivacy}. Inverse privacy refers to the situation when a user is not aware of the information that an external entity has about the user. Based on this definition, Far-reaching Insights communicate to users the inferences which can be made by the apps with superfluous permissions using advanced text and image analysis techniques. These include but are not limited to user collaboration and activity patterns; the top faces, locations, and concepts that appear in users' photos, etc.
	
	Overall, we make the following specific contributions in this paper: 
	
	\textbf{i. Far-reaching Insights sensitize users with intimate details, and promote privacy aware behavior:}
	
	Through extensive user experiments, we discover that our first simple model, \textit{Delta Permissions} fails to deter users from installing over-privileged apps. Put bluntly, \textit{telling users that their privacy is being infringed does not help.} The second model, \textit{Immediate Insights}, does twice as well in discouraging users from installing over-privileged apps. However, the clear winner is our novel model, Far-reaching Insights, which can be twice as effective in deterring users from installing over-privileged apps as Immediate Insights. 
	Overall, our analysis reveals various factors that can deter users from installing over-privileged apps. For instance, we discover that within Far-reaching Insights, \textit{Relational Insights} (that reveal users' relations with other people) reduce by half the installation of over-privileged apps, as compared to \textit{Personal Insights} (that reveal information about the \textit{users themselves}) (Section ~\ref{sec:anaUser}).

	\textbf{ii. \privyseal helps us profile developer behavior and helps users safeguard their data:}
	In an effort to promote privacy awareness in the general public, and to help users safeguard their privacy, we present \privyseal, a privacy-focused app store that uses Far-reaching Insights to warn users about over-privileged apps. This store is available for public use and has been used by over 1440 registered users. 
	A considerable fraction of these users has prior experience of using Google Drive 3rd party apps. By automatically getting their previously installed apps' metadata, we anatomize current developers' behavior, point towards potential avenues of misbehavior, and present suggestions to deter misbehavior (Sections~\ref{sec:developers} and~\ref{sec:privacyProtection}). 
	
	\textbf{iii. Shared wisdom:}
	Finally, based on our analysis we present several easy to implement practical suggestions that can be adopted by cloud providers and by those others working in the domain of privacy to safeguard users privacy in the cloud (Section~\ref{sec:privacyProtection}). 
	
	The remainder of this paper is organized as follows. Section~\ref{sec:ThirdParty} describes the 3rd party cloud app ecosystem, its threat model, and the specific case of Google Drive. In Section~\ref{sec:privacyRisk}, we describe in detail our privacy app permissions review process and results. In Section~\ref{sec:Delta}, we present our three permission models, before evaluating them in Section~\ref{sec:anaUser}. Based on the privacy-focused store we have developed (Section~\ref{sec:privysealStore}), we analyze app developer behavior in Section~\ref{sec:developers}. Finally, we give our recommendations for the community in Section~\ref{sec:privacyProtection}.

	\section{Third-party Cloud Apps Ecosystem}
	\label{sec:ThirdParty}

	There are three entities that interact in the 3rd party cloud app system: (1) a \textit{developer} who programs and manages a \textit{3rd party app}, (2) a \textit{user} who uses that app for achieving a certain service, and (3) a \textit{cloud storage provider (CSP)} hosting the user's \textit{data}. Using the CSP's API, the app gets access to a subset of the user's data after \textit{user authorization}, which is based on the user accepting a list of \textit{permissions} that determines this subset.

	\vspace{-\baselineskip}
	\subsection{Threat Model}

	Upon using 3rd party cloud apps that access their data, users sacrifice some of their privacy for getting some service(s). This tradeoff between privacy and services has been called the \emph{Privacy vs. Services Dilemma} in the literature~\cite{barnes_privacy_2006}. However, as we will see in the next section, there are many apps that require more permissions than are needed for them to function. We call such apps \textit{over-privileged apps} (as opposed to \textit{least-privileged apps} that only request the permissions needed for their functionality).
	These apps pose a \textit{risk} which can be potentially exploited, e.g., by selling data to 3rd party advertising providers. 
	A lot of the highly used 3rd party apps do not have privacy policies or justifications of the requested permissions. Moreover, the users are not usually aware of the API details or the app functionality, especially before installing the apps. Thus, the choice of installation is not well-informed from a privacy perspective. 
	In this work, we consider the 3rd party apps as the adversary (and not the CSP). We seek to combat the risk posed by over-privileged apps through improving the \textit{risk indicators} (permission models in particular) that users are presented with, during the authorization process.

	\vspace{-\baselineskip}
	\subsection{The Case of Google Drive}

	Towards that end, we have taken Google Drive as a case study, and we have anatomized this ecosystem in detail. Nevertheless, the insights gained from our analysis are applicable to other cloud platforms as well.
	To begin with, any developer can register an app that accesses Google Drive API at Google Developers Console for free. She then receives a \emph{Client ID} and \emph{Client Secret} that need to be included in the app code to access Google APIs. The developer can then specify in her code a set of Google permissions (a.k.a. scopes) she wants to obtain.
	The app itself can be hosted on any website the developer chooses; i.e., it is not hosted by Google itself. The developer can also submit a request for featuring the app on Google Chrome Web Store, which has a section for apps that work with Google Drive. In the store, apps are presented along with screenshots and descriptions of their functionality (provided by the developer). The store also allows users to rate and review apps.
	Apps can be also submitted to other web stores hosted by Google, such as the Add-ons Stores for Google Docs, Google Sheets, or Google Slides and the Google Apps Marketplace for enterprises. However, there are a lot of apps that exist outside these stores.
	
\begin{figure}[t]
		\centering
		\includegraphics[width=0.8\linewidth]{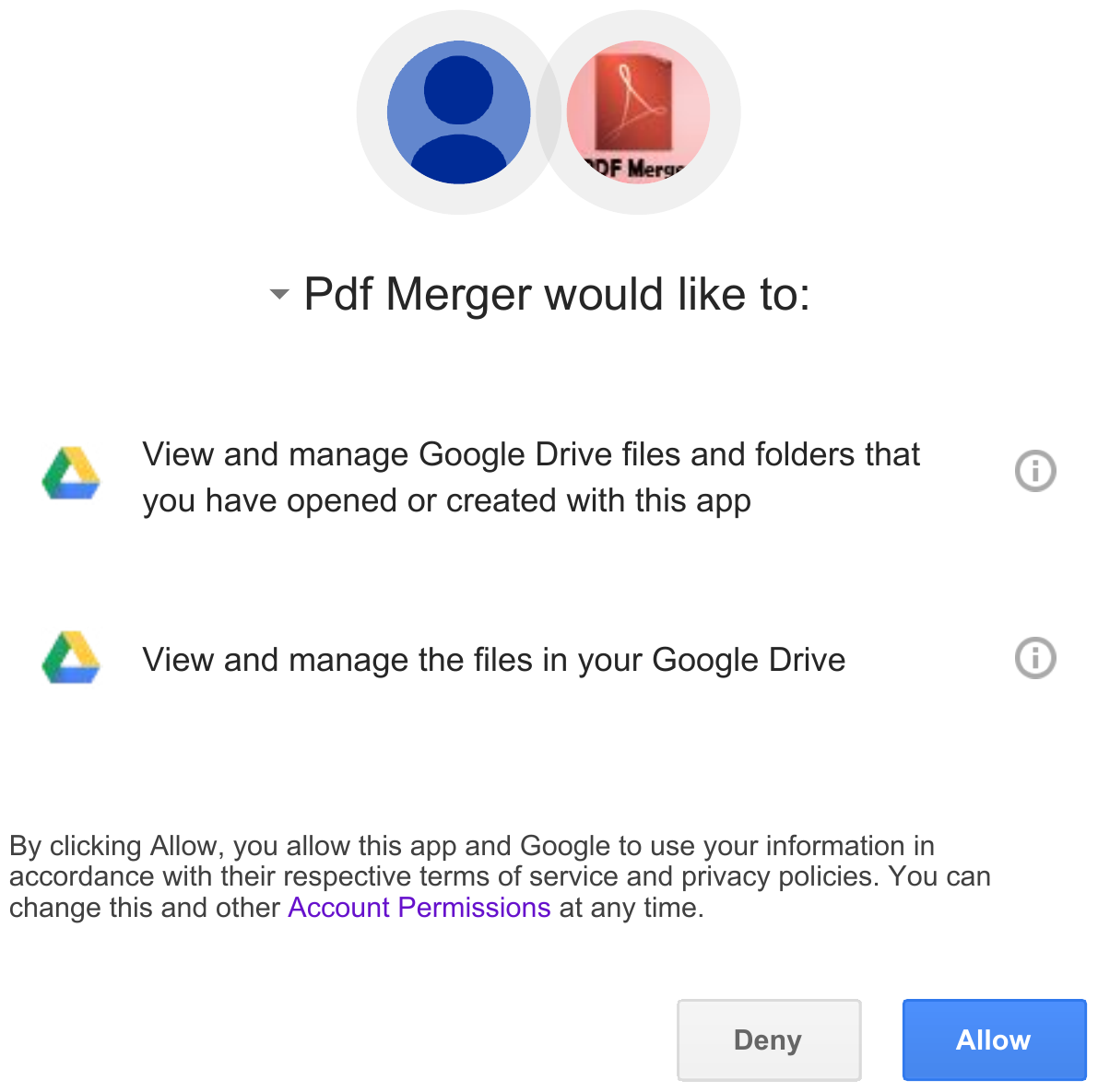}
		\caption{Example of the current permissions interface of Google Drive}
		\label{fig:baseline_pdf_converter}
	\end{figure}

	An app can request permission to access Google Drive data at any time of its operation, and not necessarily at the beginning. For example, the user can be presented with a button in a side menu that reads ``Import file from Google Drive'', and clicking on this button redirects to a Google-hosted page that presents the set of permissions requested by the app, as shown in Figure~\ref{fig:baseline_pdf_converter}. The user has to accept all these permissions to connect the app to her Google Drive. She cannot select a subset of them at installation time or later. However, she may revoke the app authorization completely from her Google account settings. As we see later, the absence of a standard location and interface for hosting apps and triggering the permissions request is one of the reasons that makes the automated, large-scale privacy analysis of apps infeasible.
	The main permissions pertinent to Google Drive are presented in Table~\ref{tab:defaultPermissions}, along with the Google-provided description for each. This short description is also presented to the user, and a longer explanation is available via clicking the info button \circled{i} next to each permission.
	
	\begin{table}
		\scriptsize
		\centering
		\begin{tabular}{L{4.5cm}|L{3cm}}
			\hline \textbf{Permission} & \textbf{Short Name} \\ 
			\hline View and manage the \textbf{files} in your Google Drive. & \textsc{drive}\\ 
			\hline View the \textbf{files} in your Google Drive. & \textsc{drive\_readonly}\\ 
			\hline View and manage \textbf{metadata} of files in your Google Drive. &  \textsc{drive\_metadata}\\ 
			\hline View \textbf{metadata} for files in your Google Drive. & \textsc{drive\_metadata\_readonly}\\ 
			\hline View and manage Google Drive files that you have \textbf{opened or created with this app}. &  \textsc{drive\_file}\\ 
			\hline View your Google Drive {apps}. & \textsc{drive\_apps\_readonly} \\ 
			\hline Add itself to Google Drive.  & \textsc{add\_drive} \\ 
			\hline View and manage its {own configuration data} in your Google Drive. & \textsc{drive\_appdata}\\ 
			\hline 
		\end{tabular}
		\newline
		\caption{Requested permissions with the short name we use for reference}
		\label{tab:defaultPermissions}
	\end{table}

\begin{figure}
		\centering
		\includegraphics[width=\linewidth]{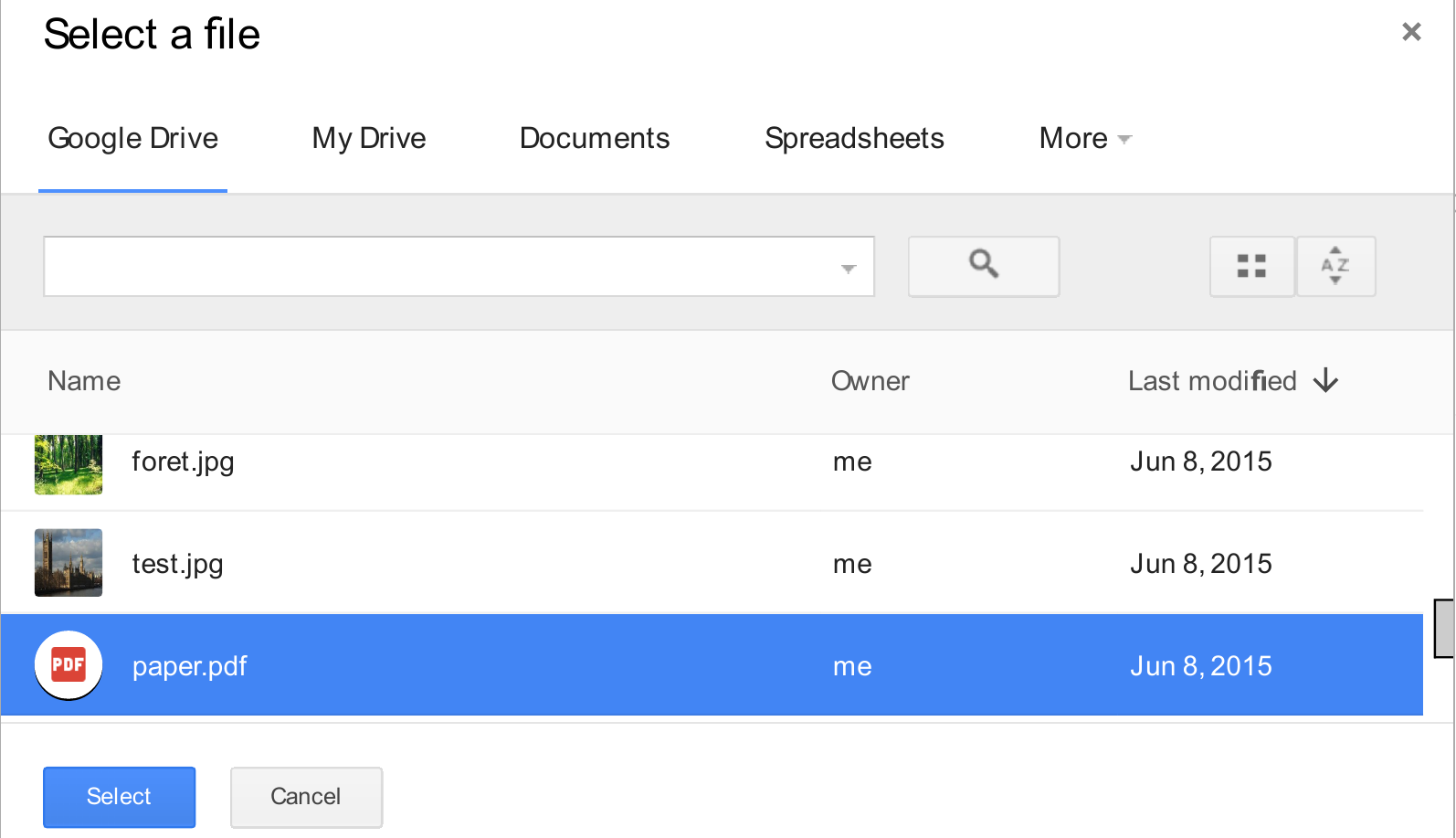}
		\caption{Google Drive file picker interface }
		\label{fig:file_picker}
	\end{figure}

	As far as files' data is concerned, an app can request access to all files (\textsc{drive} and \textsc{drive\_readonly} permissions) or on a per-file basis (\textsc{drive\_file}). When developers request \textsc{drive\_file} only, the explicit approval for each new file(s) is mediated by an interface provided by Google. For example, the developer presents the user with a Google-hosted file picker popup (Figure~\ref{fig:file_picker}) so that she can select (and thus approve access to) the file. Alternatively, the file can be opened from Google Drive's interface via the ``Open with'' option in the context menu of the file. In the case of full access, an app can access any file directly via Google Drive API without the need for user intervention. For example, this type of access enables an app to obtain all the user's files and download them in the background. 
	The developer can alternatively request access to file metadata only via \textsc{drive\_metadata} or \textsc{drive\_metadata\_readonly} (allows accessing filenames, editing dates, photos' Exif information, etc.). Additionally, the developer can request access to the list of apps the user has authorized before via \textsc{drive\_apps\_readonly}.
	It is worth noting that the permission list is not limited to Google Drive API and that it typically includes permissions from other Google APIs, such as access to user's profile information, email address, contacts list, etc.

	\section{Privacy Risk of 3rd Party Google Drive Apps}
	\label{sec:privacyRisk}

	The question that comes next is: \textbf{``what is the extent of risk that actual users are exposed to?''} To answer this, we examined a sample of 3rd party Google Drive apps to determine the percentage of apps that request extra permissions. 
	We proceeded to Google Chrome Web Store, which has a section for apps that work with Google Drive.
	The store features apps on its main page, that change with time. In total, there are around 420 apps on the store that are labeled as ``\textit{Works with Google Drive}''. We selected 100 featured apps at random from the main page (during May 2015), and we manually reviewed them one by one. Hence, our sample represents around one-fourth of the whole set of apps in the store, which is one of the main avenues for finding Google Drive apps. As we discuss later in Section~\ref{sec:anatomDevCurrent}, we discovered that, in our real word sample of 1440 users, around one-fourth of the installed apps are from the Google Chrome Store. 
	More details on the apps' dataset are presented in Appendix~\ref{apndx:reviewData}.

	\subsection{Permissions Review Process}
	\label{sec:revProcess}

	We now explain the App Permissions Review (APR) methodology we followed, and we refer the reader to Figure~\ref{fig:review} in Appendix~\ref{apndx:reviewData} for the corresponding flowchart. Our methodology is inspired by Google Drive's guide for choosing authentication scopes.
	Each APR aims to get: (a) \textbf{set $P$ of requested permissions}, (b) \textbf{set $S$ of sufficient permissions} for the app functionality. 
	We start each APR by going to the app's website, linked from the store, and testing the app manually. For each app, we first find the step where the app can be connected to Google Drive (if this is not upon the initial sign up). Then, we record the set $P$ of requested permissions and authorize the app to access a test Google Drive account created for this purpose, and we record the permissions requested. If \textsc{drive\_file} (i.e. minimal per-file access) is the only Google Drive permission requested, the app review is complete (${S=\{\textsc{drive\_file}\}}$). Otherwise, we continue to check the app's interface for all file pickers that allow importing files from Google Drive (in almost all the cases, there is at most one file picker). 
	
	In the first case where the app solely uses Google's official file picker of Figure~\ref{fig:file_picker} (e.g., an app that allows users to convert specific files to PDF format), we set ${S=\{\textsc{drive\_file}\}}$. 
	In the second case where we find that there are no file pickers in the interface and that the app functionality does not require access to any file, the app is labeled as not requiring any file permissions (${S=\{\}}$). 
	In the case where the app includes a custom file picker, we decide that (a) $S=\{\textsc{drive}\}$ if the app's declared functionality necessitates files' content (e.g., a photo collage app with custom photos browser) or (b) ${S=\{\textsc{drive\_metadata}\}}$ if the functionality does not need the content (e.g., an app that visualizes who has access to a selected folder).
	Similarly, if the app has no file picker, we decide that (a) ${S=\{\textsc{drive}\}}$ if the app's declared functionality necessitates file content (e.g., malware scanning apps for Google Drive that do not need a file picker) or (b) ${S=\{\textsc{drive\_metadata}\}}$ if the functionality necessitates file metadata (e.g., an app that visualizes all collaborators with access to user's files). 
	Finally, we label an \textbf{app as over-privileged} if either (a) the set $S$ is empty and $P$ is not or (b) if the set $P$ contains at least one permission that is more demanding than all permissions in $S$ (the five file-related permissions in Table~\ref{tab:defaultPermissions} are listed from the most demanding down to the least demanding).
	We also determine the set of \textbf{unneeded permissions $U$}, composed of each permission in $P$ that is more demanding than all permissions in $S$. The set of \textbf{needed permissions} is given by $N = P \setminus U$.

	\subsection{Review Results}

	Analyzing the APRs, we found out that \textbf{64 out of 100 apps} request unneeded permissions.
	In other words, the developers could have requested less invasive permissions with the current API provided by Google. 
	In total, 76 out of the 100 apps requested full access to the all the files in the user's Google Drive. Moreover, the 64 over-privileged apps have actually all requested full access. Accordingly, in our sample, around \textbf{84\% (64/76) of apps requesting full access} are over-privileged.
	
	\begin{table}[t]
		\centering
		\scriptsize
		\begin{tabular}
			{ |C{4cm}|c|c|}
			\hline \textbf{Permission} & \textbf{Unneeded} & \textbf{Needed}\\ 
			\hline \textsc{drive\_readonly} & 17 & 1\\ 
			\hline \textsc{drive} & 55 & 7\\ 
			\hline \textsc{drive\_metadata\_ readonly} & 2 & 1\\ 
			\hline \textsc{drive\_file} & 0 & 41\\ 
			\hline 
		\end{tabular} 
		\caption{Permissions' Usage in APRs}
		\label{tab:permissionStats}
	\end{table}

	The top permission that is needlessly requested is the full read and write access to Google Drive (in 55 apps), followed by the full read access permission (in 17 apps). This further increases the magnitude of data that can be exploited with the extra permissions.
	On the other hand, the per-file access permission is the top permission that is actually needed when requested. This happens in 41 of the apps. However, in \textbf{16 of these 41 apps}, we have found that the developer also requested full access to the user's data. Accordingly, developers are sometimes mixing full-access with partial access (which is a subset of the former).
	We note that such mixing of permissions can either be the result of developer incompetence, or it may be aimed at deceiving the user. Regardless, such apps pose a risk which can be potentially exploited. 
	Another outcome of the APR was that  \textsc{drive\_file} was the top alternative permission (in 48 apps) that could replace the unnecessary permissions requested. \textsc{drive\_metadata\_readonly} was the alternative for one app only. This indicates that, simply, the correct usage of the current Google Drive API (which does provide per-file access), can eliminate the major part of the privacy risk. Nevertheless, it is evident that developers are generally guilty of not doing this.

	\subsection{Automating the APR Process}

	Being an external party, we do not have access to the full list of Google Drive apps with their permissions. Hence, the \textit{first} task we had to do was to find the position in each app where Google Drive permissions are requested. This is not always on the main page of the app, and sometimes finding it requires navigating multiple menus and/or pages.    
	Automating this task involves building an advanced web crawler that can retrieve the permissions from a large number of such apps by smartly searching for the sign-in button. 
	The \textit{second} task was checking the functionality of the app in order to see if it matches the requested permissions. Automating the process of over-privilege detection or of real time private data leakage detection has been tackled in the mobile apps scenario (e.g. in~\cite{Felt:2011:APD} and~\cite{Enck:2010}). However, in the mobile scenario (or any similar architecture), the user's device hosts the data, the 3rd party apps, and the detection/monitoring solution. Cloud apps present a radically different scenario as the data is hosted by the CSP, the 3rd party app is served at a developer-specified location, and any detection/monitoring app would operate from outside. The only part of the code that the 3rd party app exposes is the client side code. Hence, all techniques that check the app's code (e.g., via static/dynamic analysis) or its inputs/outputs cannot be transplanted to the cloud app case as they would evidently underestimate what APIs/permissions the app might need. 
	The only automatic way we perceive for over-privilege detection is to cluster apps of similar functionality and identify the ones which request more permissions than others in the same cluster. Even then, the data collected manually would be used as the ground-truth to evaluate the automated method. Detecting actual data leakage is much more challenging in the cloud apps scenario as the app can send users' data to other parties from the server side (which is impossible to monitor via external solutions).
	Faced with these limitations, manual expert reviews are the closest we can get to assessing the apps' needed permissions. Still, we do not claim that this method is perfectly accurate as a developer might be working on a non-advertised feature that requires new permissions. However, we conjecture that APRs are accurate with the vast majority of the reviewed apps\footnote{From our experience over one year, rarely did apps introduce new features that required new permissions. Moreover, in Section~\ref{sec:privysealStore}, we discuss how to further alleviate this concern in a real-world deployment by allowing developers to submit rebuttals.}.
	Finally, as our main purpose in this work is to characterize the ecosystem and suggest alternative permission models, automating both the app permissions collection and the over-privilege detection tasks falls out of the scope of this work. We note though that we are concurrently working on the specific research problem of designing automated APRs.

	\section{New Permission Models}
	\label{sec:Delta}

	In the light of the risk that over-privileged apps pose, we propose in this section three alternatives to the existing permission model in Google Drive before evaluating their efficacy in mitigating the risk in the next section.

	\subsection{Delta Permissions}
	
	Our first model is based on the following hypothesis:
	\textit{``When users are informed about the unneeded permissions being requested by apps, they are less likely to authorize such apps.''}
	\begin{figure}[t]
		\centering
		\includegraphics[width=0.8\linewidth]{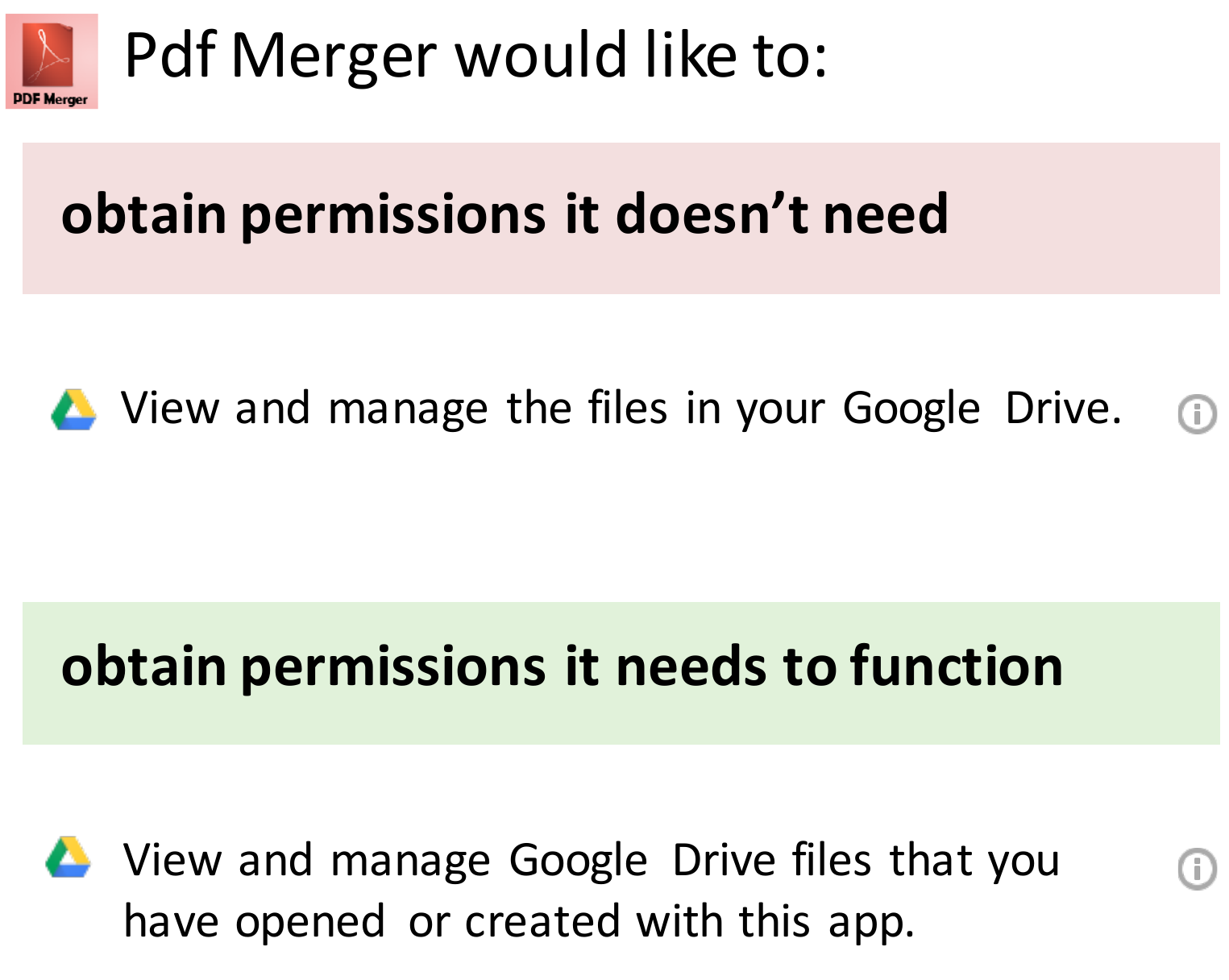}
		\caption{Example of Delta Permissions interface}
		\label{fig:delta_example}
	\end{figure}
	Hence, this model replaces the current permissions interface displayed in Figure~\ref{fig:baseline_pdf_converter} with a new interface, presented in Figure~\ref{fig:delta_example}. We call this permission model \textit{Delta Permissions (DP)}, and it reveals to the user the distinction between permissions that are needed for the app functionality and those others (the delta) that are unnecessarily requested.

	\subsection{Immediate Insights}
	\label{sec:immediateInsights}
	
\begin{figure*}[t!] 
		\centering
		\minipage{0.73\textwidth} 
		\fbox{\includegraphics[width=\linewidth]{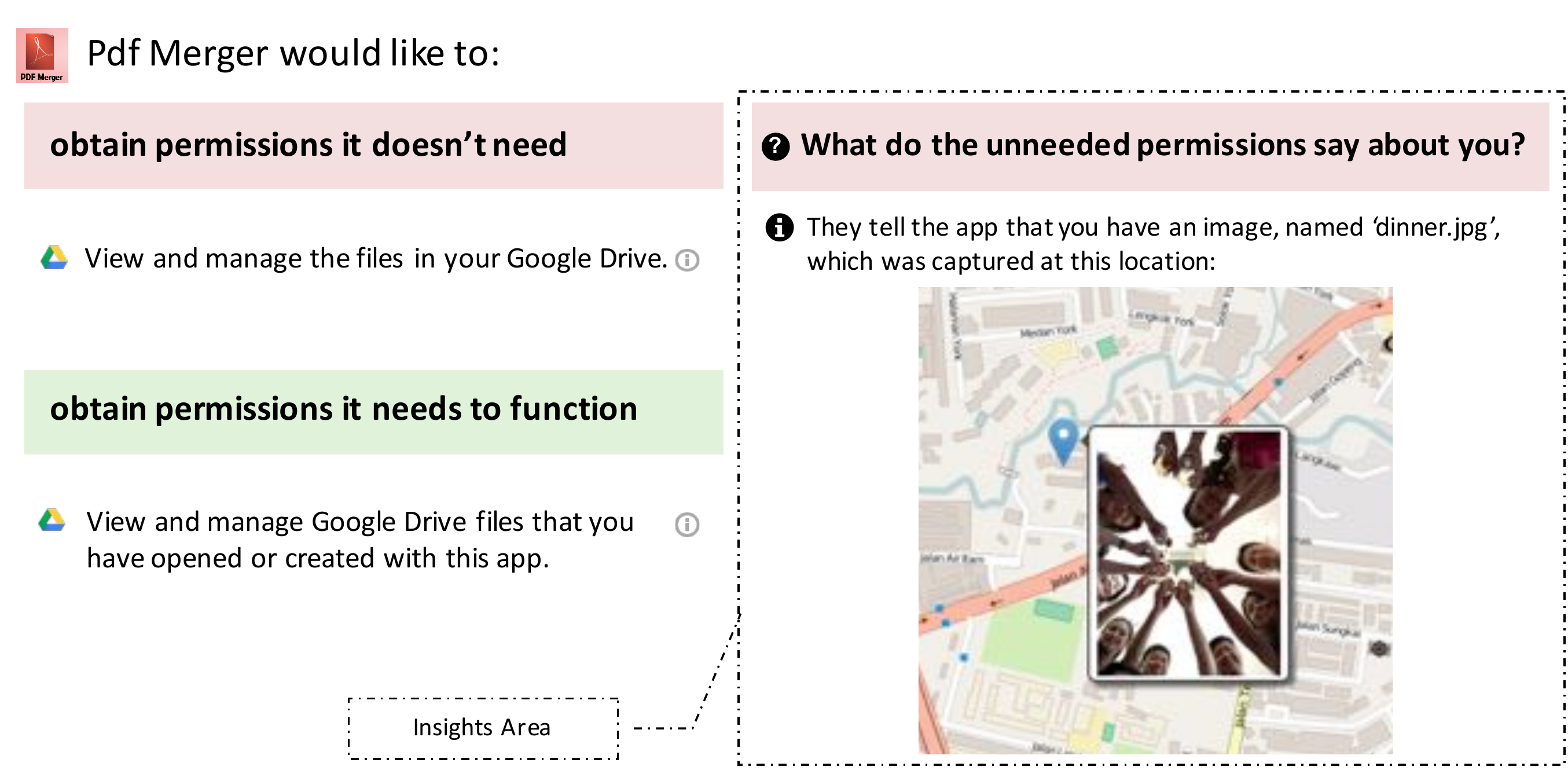}}
		\caption{Example of Immediate Insights interface; the same layout is used for Far-reaching insights, with the insights area content changing accordingly}
		\label{fig:personalizedInsights}
		\endminipage\quad
		\minipage{0.23\textwidth} 
		\fbox{\includegraphics[width=\linewidth]{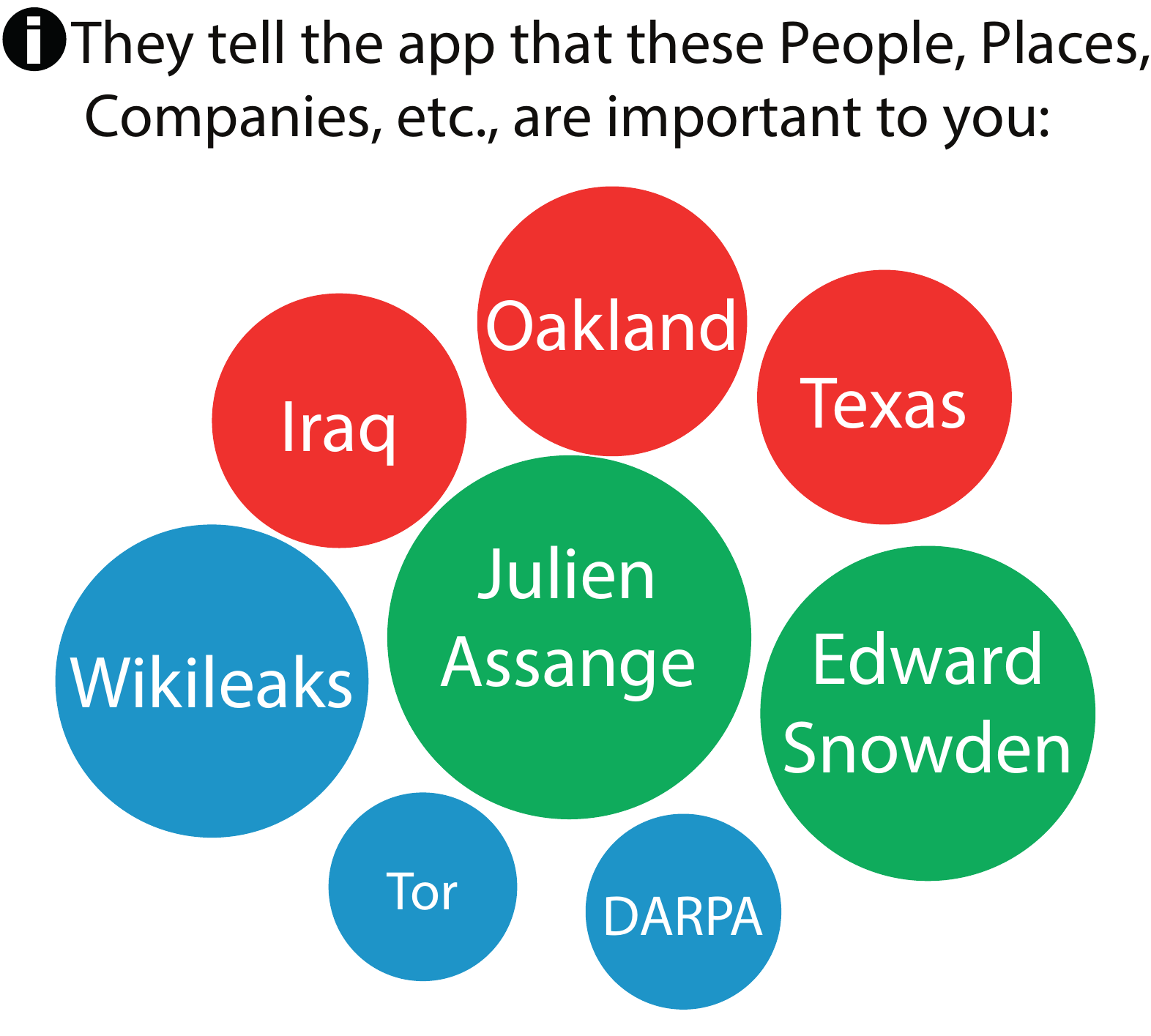}}
		\caption{Named \textit{Entities} insight}
		\label{fig:entities}
		\endminipage
	\end{figure*}

	The second model is based on the following hypothesis:
	\textit{``When users are shown samples of the data that can be extracted from the unneeded permissions granted to apps, they are less likely to authorize these apps.''}
	Accordingly, we show users randomly selected data examples, directly extracted from their Google Drive, such as excerpts of text or image files, photo locations, or people she collaborated with.
	An instance of this model, which we call \textit{Immediate Insights (IM)}, is given in Figure~\ref{fig:personalizedInsights}. On the left, we have the same previous \textit{DP} interface. On the right, we have the \textit{Insights Area}, where we show a question that says: ``What do the \textbf{unneeded permissions} say about you?'', followed by an answer in the form of a visual with short explanatory text. In this figure, the Insights Area visualizes the location where a randomly chosen user photo was taken.
	In the following, we describe the design of the \textit{IM} Insights:
	
	\tparagraph{Image}
	We show an image selected at random from the set of user's image files.
	
	\tparagraph{Location}
	We randomly choose a photo from the set of user's image files, such that it includes a GPS location in its Exif data. Then we show that photo on a map centered at that location (as in Figure~\ref{fig:personalizedInsights}).
	
	\tparagraph{Text}
	We show the user an excerpt from the beginning of a randomly chosen textual file.
	
	\tparagraph{Collaborator}
	We show the profile picture and the name of a randomly chosen collaborator.

	\begin{figure*}[t!] 
		\centering
		\minipage{0.26\textwidth} 
		\fbox{\includegraphics[width=\linewidth]{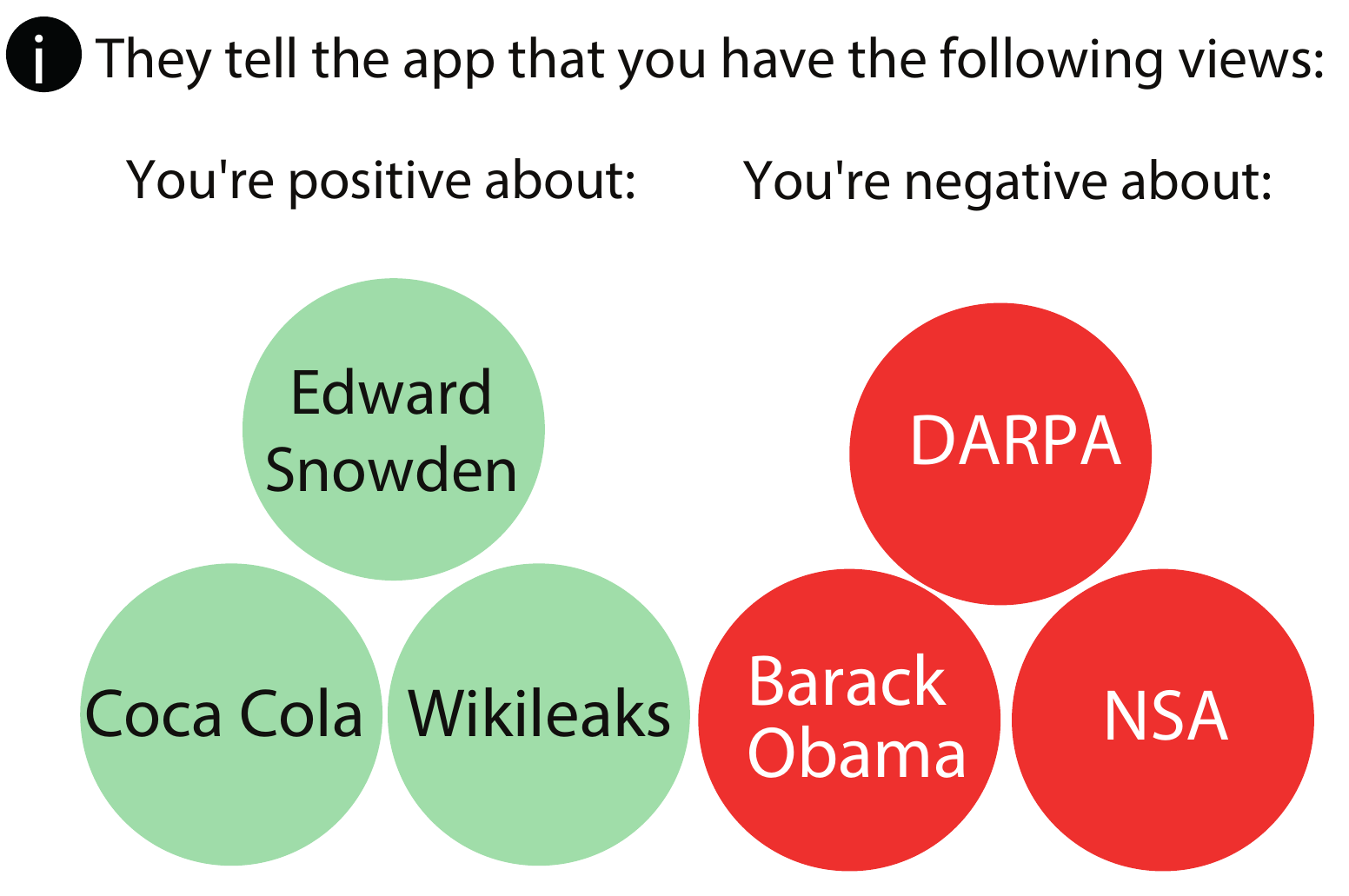}}
		\caption{\textit{Sentiments} insight}
		\label{fig:sentiments}
		\endminipage\hfill
		\minipage{0.24\textwidth} 
		\fbox{\includegraphics[width=\linewidth]{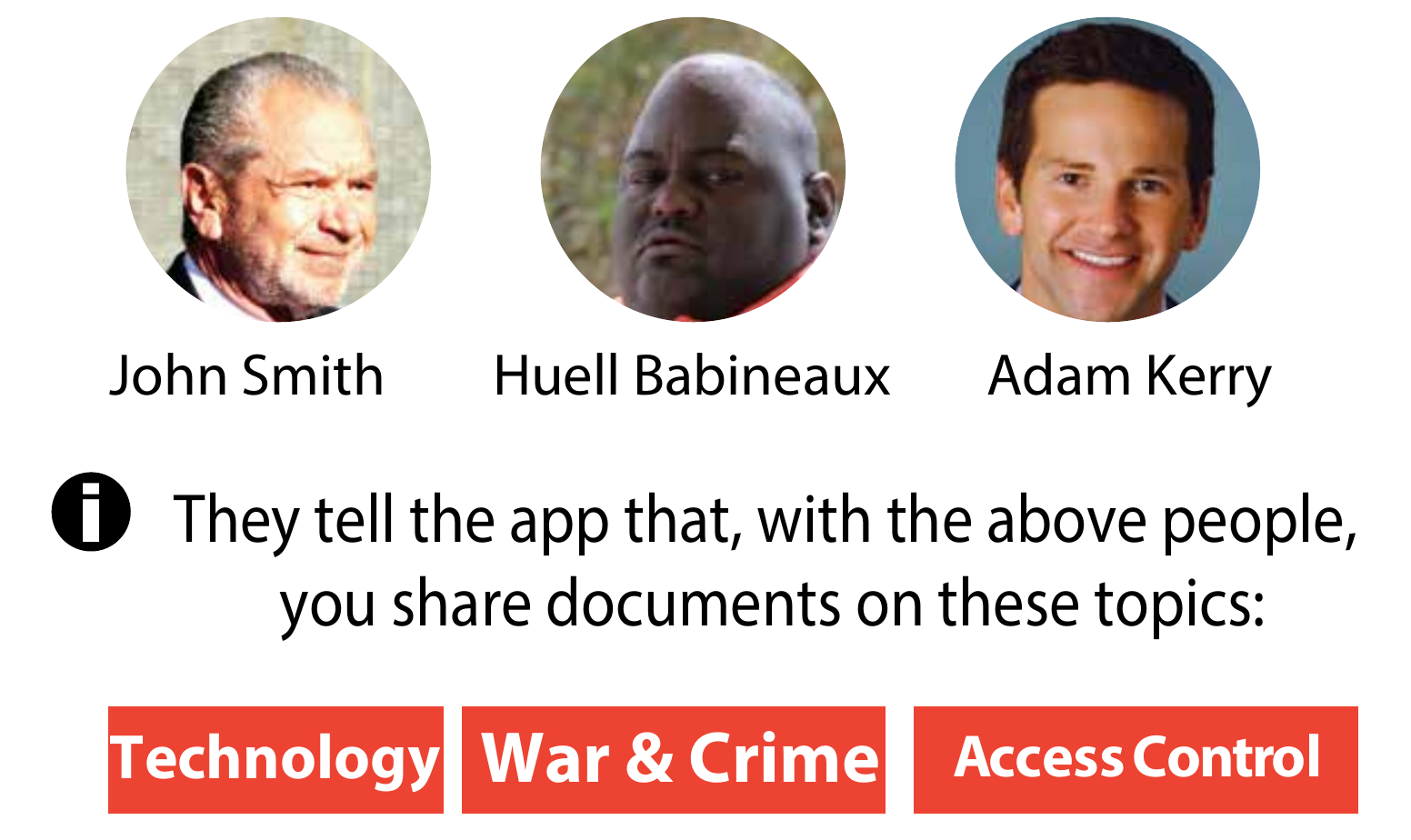}}
		\caption{\textit{SharedInterests} insight}
		\label{fig:sharedInterests}
		\endminipage\hfill
		\minipage{0.23\textwidth} 
		\fbox{\includegraphics[width=\linewidth]{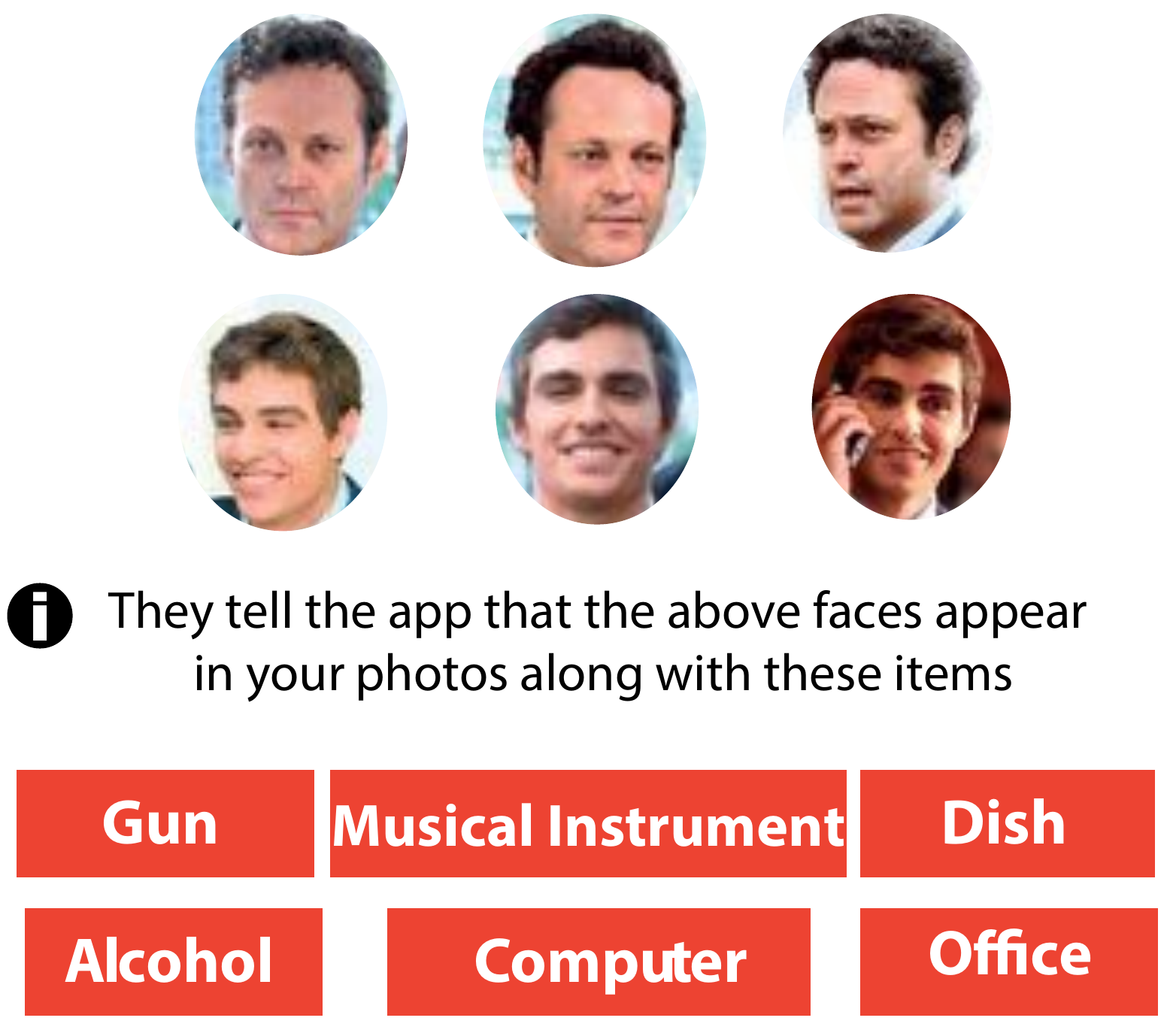}}
		\caption{\textit{FacesWithContext} insight}
		\label{fig:faceswithcontext}
		\endminipage\hfill
		\minipage{0.22\textwidth}
		\fbox{\includegraphics[width=\linewidth]{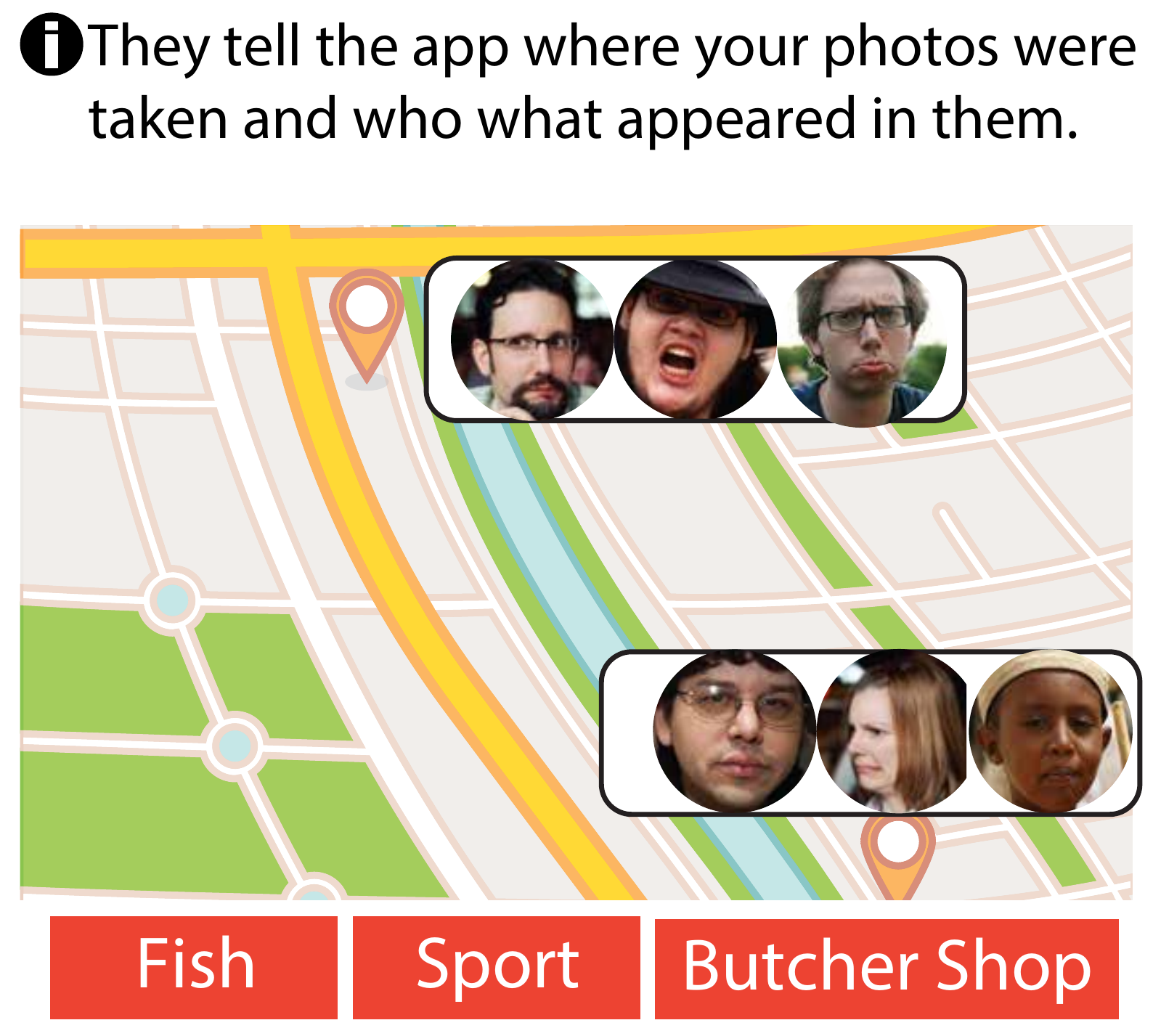}}
		\caption{\textit{FacesOnMap} insight}
		\label{fig:facesonmap}
		\endminipage 
	\end{figure*}
	
	\subsection{Far-reaching Insights} 
	\label{sec:farInsights}
	
	The third model is based on the following hypothesis:
	\textit{``When the users are shown the far-reaching information that can be inferred from the unneeded permissions granted to apps, they are less likely to authorize these apps.''}
	These are insights that go beyond examples and include what can be inferred by running more involved algorithms, such as sentiments towards entities, objects identified in photos, faces detected, etc. Hence, we denote this model by \textit{Far-reaching Insights}  (or shortly \textit{FR Insights}).
	The interface layout is the same as that of Figure~\ref{fig:personalizedInsights}, but with the Insights Area containing an \textit{FR} insight instead of an immediate insight.
	In this work, we have designed 6 types of \textit{FR} insights that can be extracted from users' data. Below, we briefly describe each of them and we refer the reader to Appendix~\ref{apndx:Insights} for the algorithms used for their generation:\\
	
	\tparagraph{Entities, Concepts, and Topics (ECT)}
	The first type of insights we form is based on applying various NLP techniques to extract \textit{named Entities (E)}, \textit{Concepts (C)}, and \textit{Topics (T)} from users' textual files. Entities, extracted via \textit{Named Entity Recognition}, might include people the user works with, companies she talks about, places she plans to visit, etc. 
	Concepts are extracted tags from the user's documents. For example, the sentence ``My favorite brands are BMW, Ferrari, and Porsche'', would be tagged by the concept ``Automotive Industry''. Topics are higher level abstractions (e.g. technology, art, business, etc.) that can be used for classifying users' documents. Both concepts and topics can serve for profiling users' interests. We combine these together due to the similar nature of these insights. When we use \textit{ECT}, one of $E$, $T$ or $C$ is randomly displayed to the user in the Insights Area.
	For purposes of text analysis in this work, we partially used AlchemyAPI's service with random excerpts of text extracted from users' documents. Users in our experiments were informed about this on the main page of the web app they sign in to.

	\tparagraph{Sentiments}
	We used sentiment analysis in order to identify the entities with the most positive or negative sentiments and then display them to the user in the Insights Area as in Figure~\ref{fig:sentiments}.

	\tparagraph{Top Collaborators}
	The next insight we added displays the top collaborators a user has, based on the analyzed files. These typically include close work colleagues, intimate friends, or people the user goes out with and shares pictures with afterward.

	\tparagraph{Shared Interests}
	In this insight, we represent the user's mutual topics of interests with a group of people. As shown in Figure~\ref{fig:sharedInterests}, the Insights Area would then contain a list of people alongside different topics of documents shared with these people.
	
	\tparagraph{Faces with Context}
	We now come to the insights that are based on features inside the user's images. The first insight of this type shows a group of faces, representing the most frequent people appearing in the user's images, alongside the concepts that appear in the same images (see Figure~\ref{fig:faceswithcontext}). One can imagine that such information might be valuable, for example, to advertisers that aim to extract the user's interests in certain people, products, or services.
	
	\tparagraph{Faces on Map}
	In addition to the image content itself, image metadata can be also sensitive, especially the geographical location where the image is captured. Hence, this insight, shown in Figure~\ref{fig:facesonmap}, relays to the user the places where her photos are taken, in addition to the faces and items in those photos.

\begin{figure}[t]
		\centering	\includegraphics[width=0.7\linewidth]{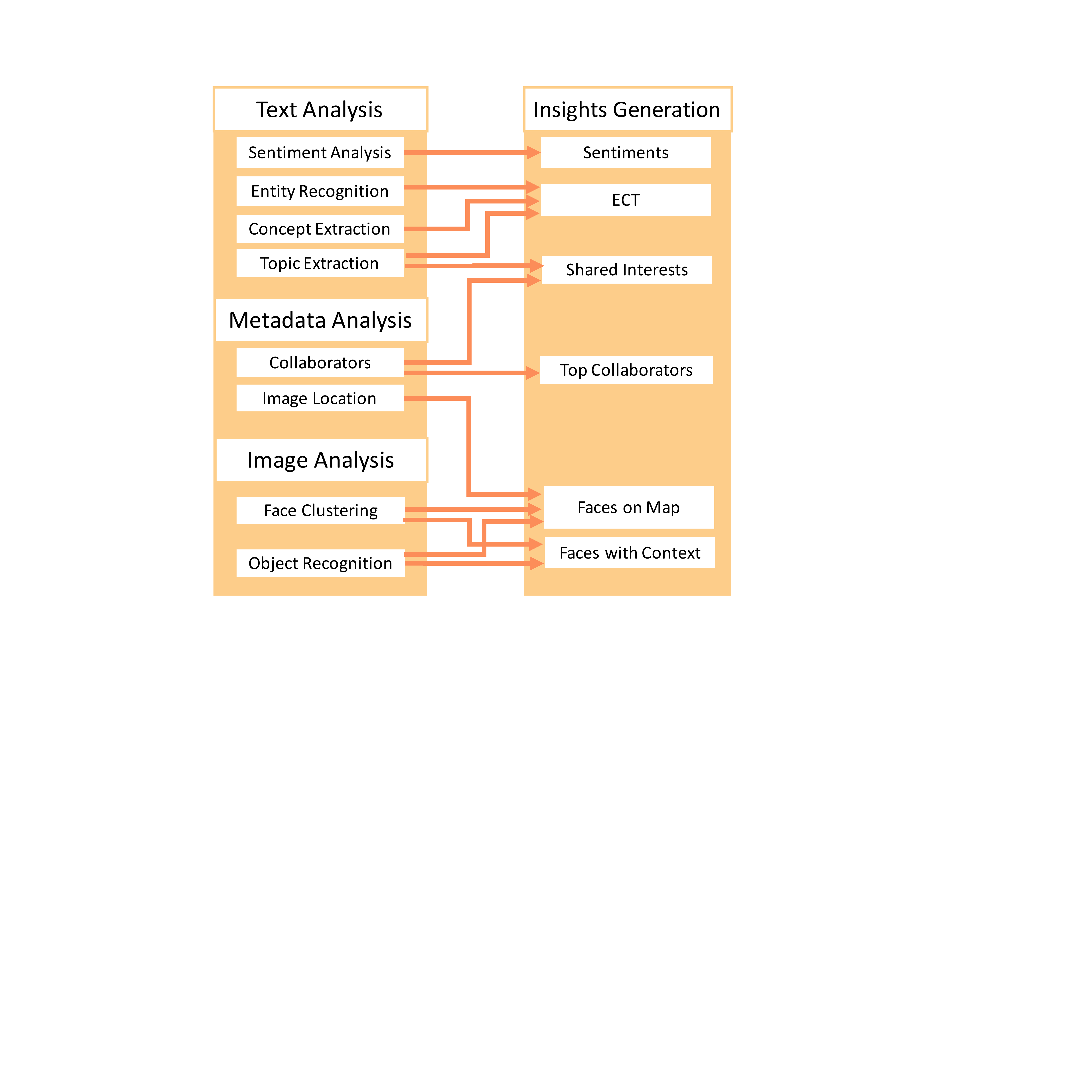}
		\caption{Component diagram mapping the used analysis techniques to the generated insights}
		\label{fig:insights_generation}
	\end{figure}

	\noindent The component diagram of Figure~\ref{fig:insights_generation} summarizes the techniques used for generating each of the \textit{FR} insights.
	
	\tparagraph{\textit{Further Notes}}
	We note that the reasoning behind designing lightweight models such as \textit{DP} and \textit{IM} was that we wanted to examine whether designing heavyweight insights such as \textit{FR} Insights is worth the effort for us and the potential adopters of our approach, or do users respond equally favorably (or badly) to both the heavy and the lightweight approaches, in which case \textit{FR} insights need not be adopted? 
	We also note that, for an app that does not request unneeded permissions (even if it requests full access), the Insights Area will simply show a text saying that the app doesn't require any extra permissions.
	We also note that we follow Google Drive's approach of requesting permissions ``At Setup''~\cite{schaub2015design} (i.e., at the first time of app authorization).
	This is unlike other ecosystems (e.g., iOS or Android M), which require a ``Just in Time'' approach (i.e., permissions are requested only when the actual functionality is needed).
	This is because, in Google Drive, many apps are supposed to work with the user's data even when she is offline. Hence, granting access in an interactive manner for individual permissions is not always feasible.

	\section{Evaluating the Models}
	\label{sec:anaUser}

	\subsection{Experimental Setup}
	
	\label{sec:Exp1}
	
	We designed an experiment with actual users in order to test the hypothesis of whether the new models can better deter the users from installing over-privileged apps as compared to the existing one, and to discover factors that influence users' decisions. 
	\minisec{User Recruitment}
	In order to recruit users, we primarily used our university's mailing list.
	The users were briefed about an app that is related to protecting the privacy of their data against 3rd party apps on Google Drive. The news about the app was also reported on the university's website and was picked up by several technology websites.
	The website described itself as an app for Google Drive that aims at exposing what 3rd party web apps can needlessly get about users. 
	
	Via our website, the users can sign in to their Google account and then grant full Google Drive access to our app. 
	Next to the ``sign in'' button, we linked to our privacy policy, explaining what data the app gets and what it keeps. 
	Only those users who had at least 10 files containing text or 20 images were allowed to continue. This is to ensure that they possess at least a minimal level of experience with Google Drive. Figure~\ref{fig:userFileStats} of Appendix~\ref{appndx:experiment} shows the density plot of the percentages of users' analyzed files that are textual. Next, users who agreed to participate in our experiment were randomly assigned to one of the groups described below. As a motivation to complete the experiments, the users were enrolled in a lucky draw, where they could win one of five gift cards to a mobile app store of their choice.
	
	\vspace{-\baselineskip}
	\minisec{Methodology}
	
	The first goal of the experiment is to investigate the efficacy of the three new permission models by comparing them to the existing Google Drive permission model as well as to each other. The second goal is to perform micro-comparisons among the different types of \textit{IM} and \textit{FR} insights. Accordingly, we went for a mixed between-subject and within-subject design. The reason for not going for a complete between-subject design was the large number of participants needed for statistically significant results with 12 independent groups (for all micro-comparisons). The reason for not going for a complete within-subject design was to avoid any participants' bias that can result from showing the existing interface they are used to and the new interface we developed in the same experiment. 
	Accordingly, we had four groups in our experiment. A user is assigned to one single group, and the only permission interface she sees during the experiment is that of its group. The groups were:
	
	\begin{enumerate}
		\item
		\textbf{Baseline group (\textit{BL} group):} Users in this group were presented with a clone of the original interface that Google shows upon installing the app (shown in Figure~\ref{fig:baseline_pdf_converter}). This group serves as the control group, and we briefly refer to it as \textit{BL} 
		\item
		\textbf{Delta Permissions group (\textit{DP} group):} Users in this group were presented with the modified interface, previously shown in Figure~\ref{fig:delta_example}.
		\item
		\textbf{Immediate Insights Group (\textit{IM} group):} Users in this group were presented with the modified interface of Figure~\ref{fig:personalizedInsights}, with the Insights area containing one of the \textit{IM} insights of Section~\ref{sec:immediateInsights}.
		\item
		\textbf{Far-reaching Insights Group (\textit{FR} group):} Users in this group were presented with the modified interface, of Figure~\ref{fig:personalizedInsights}, with the Insights Area containing one the \textit{FR} insights described in Section~\ref{sec:farInsights}.
	\end{enumerate}

	A user experiment was divided into multiple \textit{tasks}. 
	In each task, the user was requested to select an app with a specified \textit{goal}.
	For example, the goal would read ``Select the app which allows you to extract the ZIP files on your Google Drive'', and the corresponding app would be ``ZIP Extractor''. 
	The user would then choose this app among other apps that are listed in the interface. We show this interface in Figure~\ref{fig:task} of Appendix~\ref{appndx:experiment}, and we note that it is similar to the actual Google Chrome Web Store. 
	Moreover, only one app of those listed satisfies the given goal, and it is highlighted in the interface.
	This part of the setup only serves a gamification purpose in order to keep the user interested. 
	Once the user selects the app, she is presented with a permission interface that corresponds to its group (i.e. that of Figure~\ref{fig:baseline_pdf_converter} for the \textit{BL} group, Figure~\ref{fig:delta_example} for the \textit{DP} group, and Figure~\ref{fig:personalizedInsights} with a randomly selected visual for the \textit{IM} or \textit{FR} Insights groups). 
	The user is then presented with a question that says: ``Based on permissions below, would you be likely to install this app?''. She can choose between ``Permissions are too invasive'' (accept) and ``I'm OK with these permissions'' (reject). 
	We worded the question so that we avoid all users rejecting the installations of all apps. We rather aimed that users would reject apps whose permissions they consider as too invasive, thus allowing us to do within-subject comparisons.
	After answering the question, the user is directed to the next task with another app, until she completes the whole set of tasks. 
	
	The apps used in the experiment were obtained from the Google Drive section of the Chrome Web Store. For experimental purposes, we modified the permissions requested by these apps to be able to test various conditions. 
	Unlike in the store, we removed elements such as ratings, user reviews, and screenshots and kept a minimal interface, allowing the users to focus solely on the app permissions. 
	We also avoided using apps from popular vendors to avoid the bias resulting from users being influenced by famous brands.
	These steps were taken to study the effect that the permission model has on the user's decisions, without the influence of extraneous factors\footnote{Incidentally, the user might confront a scenario exactly as in the experiments if she does not find the app from the store, but lands on a certain site that has the option of authenticating with Google Drive.}. Moreover, the apps were presented to the users in randomized order to compensate for the effects of learning and fatigue.
	For reference, the permissions that each app requested are presented in Table~\ref{tab:experimentPermissions}.
	A user assigned to the \textit{BL} or \textit{DP} groups had to install 5 apps in 5 tasks. For the \textit{IM} Insights and \textit{FR} groups, we added additional apps. This is because we wanted to compare the effects of the different kinds of insights. The permissions of the additional apps were fixed to those of (ZIP Extractor), but the insights displayed were changing. For each user, the insights were assigned at random to each of those added apps. In total, users assigned to the \textit{IM} Insights and \textit{FR} Insights groups had to complete 8 and 10 tasks respectively.

	\begin{table}[t]
		\scriptsize
		\centering
		\begin{tabular}{L{2cm}| C{0.5cm} C{1.2cm} C{0.8cm} C{0.9cm}} 
			\hline App   & \textsc{drive} & \textsc{drive\_ metadata}& \textsc{drive\_ file} & group\\
			\hline ZIP Extractor & R,U & & R,N & 1,2,3,4\\
			\hline Xodo \textsc{pdf} Viewer \& Editor & &  R,U & R,N & 1,2,3,4\\
			\hline WhoHasAccess & R,U & R,N &  & 1,2,3,4\\
			\hline Video Converter & R,U  & &  & 1,2,3,4\\
			\hline Cloud Convert &   R,U  & & NR,N & 1,2,3,4\\
			\hline HelloFax &   R,U  & & R,N&  3,4\\
			\hline Heap Note   & R,U  & & R,N&  3,4\\
			\hline Photo to\\ Cartoon  & R,U  & & R,N&  3,4\\
			\hline PDFUnlock &   R,U  & & R,N&  4\\
			\hline HelloSign   & R,U  & & R,N&  4\\			
			\hline
		\end{tabular}\caption{Permissions of apps in the experiment. \textit{R}: Requested, \textit{N}: Needed, \textit{U}: Unneeded, \textit{NR}: Not Requested}
		\label{tab:experimentPermissions}
	\end{table}

	\minisec{Data Protection and Ethics}
	Respecting the user privacy when working with cloud data is of fundamental importance. Our experiments were done according to a code of ethics protecting this privacy. In particular, after generating the insights from a user's files, these files are deleted immediately from our apps' servers. As per our displayed privacy policy, only the insights' data presented to the user is kept in the app database. Moreover, the user is given the option to delete her insights data at any time with a single click in the app's menu. The database dump we ran our analysis on was isolated from the one to which the deployed web server connects. Also, we use the \textit{https} protocol so that users can securely connect to our system. Before data analysis, we anonymized any occurrence of names and emails in the database by applying a one-way MD5 hash on them. At all times, we refrained from manually checking the database for users' insights.
	All the images used in this paper are in the public domain, and the insights shown do not belong to real users. For further transparency, all the libraries and frameworks used for building the tool and data analysis were listed and linked to from the main page of the website. Although an IRB review was not performed beforehand, this paper was subsequently reviewed by our university’s IRB, which did not object to publishing the results.

	\subsection{Results}
	\label{sec:reviewResults}
	
	We got 210 users in total who successfully completed this part of the experiment. Out of them, 55 were in the \textit{BL} Group, 50 in the \textit{DP} Group, 54 in the \textit{IM} Insights group, and 51 in the \textit{FR} group. 
	We start by interpreting the results of our user experiment and comparing the efficacy of the various permission models.
	The metric we used in our evaluation is the \textit{Acceptance Likelihood $AL$}, defined as:
	\begin{equation}
		AL= \frac{ \# \mbox{(\textit{Accepts}) } }{ \# \mbox{(\textit{Accepts})} + \#\mbox{(\textit{Rejects})} },
	\end{equation}
	where \textit{Accepts} denotes the cases where users were fine with the permissions, and \textit{Rejects} denotes the cases where they found them too invasive. \textit{Accepts} and \textit{Rejects} are aggregated across users and tasks for the permission model under consideration. The lower the $AL$, the better the performance in deterring users from installing over-privileged apps.

	\begin{figure}[t!] 
		\minipage{0.5\textwidth} 
		\includegraphics[width=.95\linewidth]{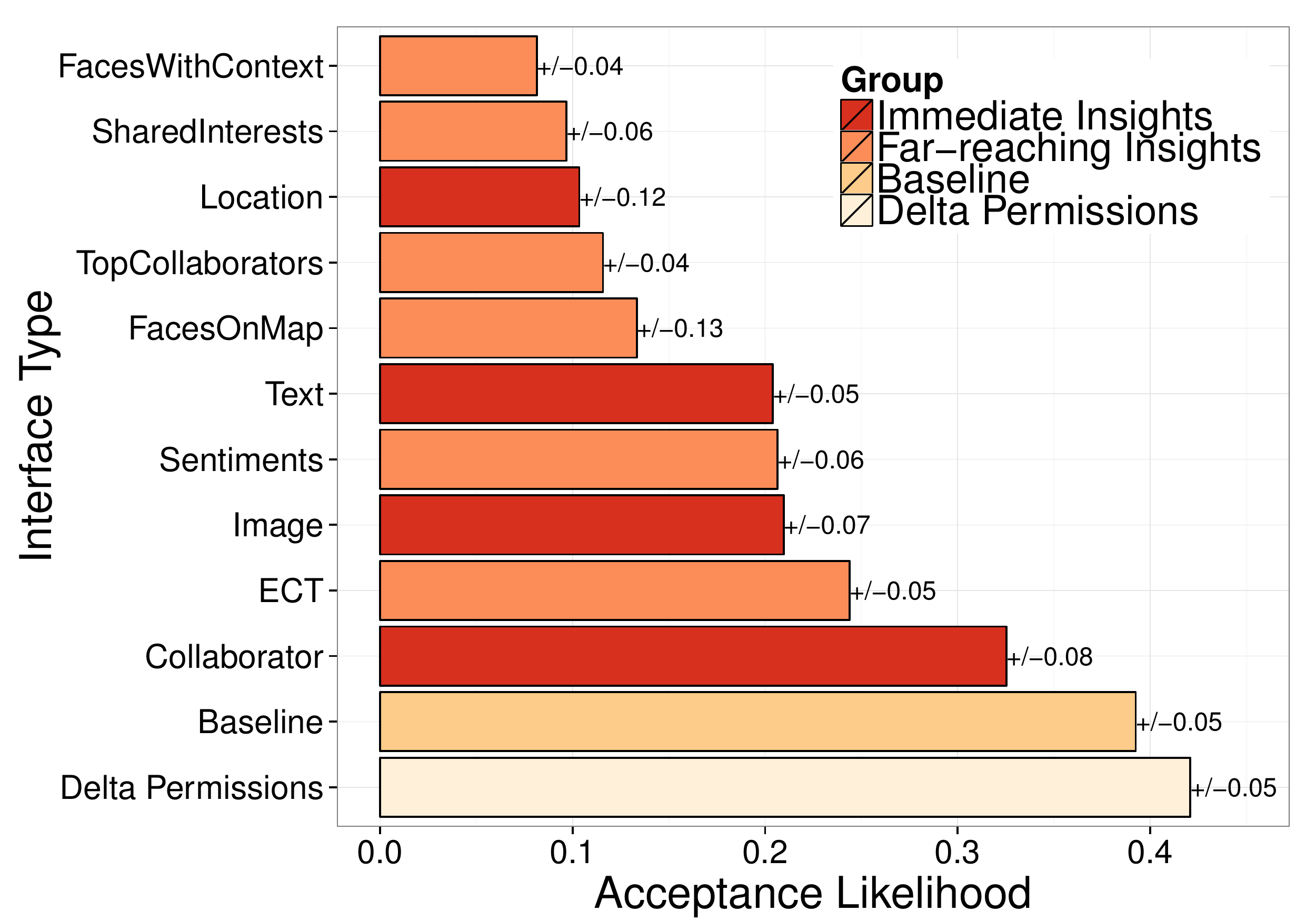}
		\centering
		\caption{$AL$ for the different types of interfaces; numbers next to each bar are the error values at 95\% confidence interval}
		\label{fig:visuals_comparison}
		\endminipage\hfill
		\minipage{0.5\textwidth}
		\centering
		\includegraphics[width=0.9\linewidth]{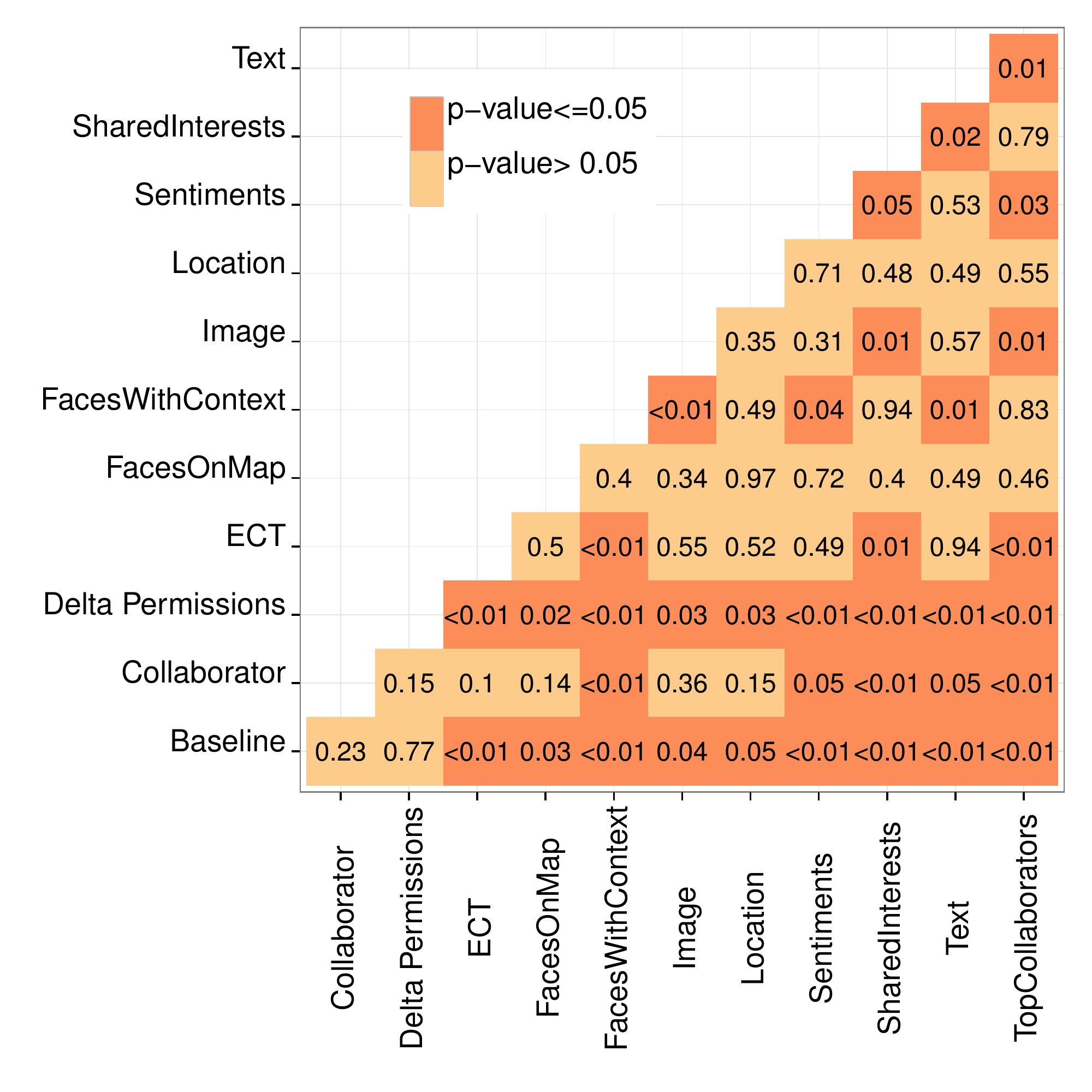}
		\caption{$p$\mbox{-values} of pairwise tests; if $p$\mbox{-value}$\le0.05$ we consider the visuals on the corresponding row and column as different; the difference direction is obtained from Figure~\ref{fig:visuals_comparison}}
		\label{fig:pvaluesTable}
		\endminipage 
		
	\end{figure}

	In order to compare the effect of different interfaces, we plotted in Figure~\ref{fig:visuals_comparison} the Acceptance Likelihood for the \textit{BL} and \textit{DP} groups and also for each particular insight of the \textit{IM} and \textit{FR} Insights groups.
	To evaluate the significance of the $AL$ differences among the interface types, we fit a generalized linear mixed model (GLMM) with the user's decision (Accepting/Rejecting the app installation) as the binary response variable and the interface type as the fixed effect. Participants' IDs and apps' names were fitted as random effects to control for the potential between-participants and between-apps variabilities. The model was fit assuming a binomial distribution and a logit link function, using the glmer function in the \textit{lme4} package in \textit{R}~\cite{lme4}. Visual inspection of residual plots did not reveal any obvious deviations from homoscedasticity or normality. The significance of differences among $AL$ values was determined using Tukey Honest Significant Difference test with the \textit{glht} function of the \textit{multcomp} package~\cite{multicomp}. The difference between the $AL$ of any two interfaces in Figure~\ref{fig:visuals_comparison} is significant if the corresponding row and column intersect at a  $p$\mbox{-value}$\le0.05$ in Figure~\ref{fig:pvaluesTable}.

	\tparagraph{Inefficacy of Baseline and Delta Permissions}
	We first found that the Delta Permissions and Baseline approaches performed closely ($AL$ of 0.42 and 0.39 respectively) without a statistically significant difference ($p$\mbox{-values}$=0.77$). Hence, we found no evidence of any advantage that the \textit{DP} can introduce, which means that telling our experiment's participants explicitly about unneeded permissions did not help deter them from installing over-privileged apps. We also observe that both these interfaces had a significantly higher $AL$ (i.e. $p$\mbox{-values}$\le 0.05$) than all the insights, except for the Collaborator insight. This highlights the fact that showing well-selected insights will result in deterring more users compared to the case of not showing any insights.

	\tparagraph{The Power of Relational Insights}
	The next interesting outcome is that there is a category of insights (Category 1) composed of \{\textit{Image}, \textit{Text}, \textit{ECT}, and \textit{Sentiments}\} that are all associated with a significantly higher acceptance likelihood than the category composed of \{\textit{FacesWithContext}, \textit{TopCollaborators}, and \textit{SharedInterests}\} (Category 2)\footnote{The number of users who had location-tagged photos was low; hence, we could not obtain highly significant results in the case of \textit{Location} and \textit{FacesOnMap} insights. }. 
	Since this is a very interesting result, we investigate further to analyze the defining characteristics of these two naturally clustered categories. The main feature of Category 1, which includes both \textit{IM} and \textit{FR} insights, is that insights in this category are restricted to characterizing the user \textit{herself}, such as showing text excerpts from her documents, topics appearing in them, or images she has in her files. Hence, we denote this category as \textit{Personal Insights}.
	On the other hand, the defining feature of Category 2 insights, which are all Far-reaching, is that they extend to characterizing the relationships of the user \textit{with other people}. For instance, \textit{FacesWithContext} shows the most important faces in user's photos along with the items appearing with them. \textit{SharedInterests} shows the people who collaborate with the user and the type of topics they share. Also, \textit{TopCollaborators} identifies the most frequent people the user interacts with. We denote these as \textit{Relational Insights}. From our results, we can conclude that Relational Insights promote greater privacy awareness in users, as such insights are more likely to dissuade them from installing over-privileged apps.
	
	\tparagraph{Impact of Face Recognition}
	Delving deeper into more results brought forth by the comparison of different insights, one can notice that showing examples of user's images ($AL=0.21$) is significantly less deterring than showing the important faces and listing the concepts in the image ($AL=0.08$) with pairwise comparison $p$\mbox{-value}$ <0.01$. This highlights the fact that users are sensitive towards the output of face detection and object recognition in photos. Given that services such as Google Photos, OneDrive, and Flickr already apply such techniques to facilitate search, the above result highlights that they can also be used by these companies to easily implement solutions such as ours for raising users' privacy awareness when sharing data.
	
	\tparagraph{Influence of High-Level Textual Insights}
	Contrary to the case of images, in the case of textual documents, showing the high-level entities or concepts extracted from the text does not seem to have a significant difference as compared to simply showing direct excerpts from the text ($p$\mbox{-value}$=0.94$). Only when the relationship factor is introduced does the $AL$ significantly decrease (as in the case of \textit{SharedInterests}).

	\tparagraph{Superiority of Far-reaching Insights}
	By aggregating the results over all the experiments with \textit{FR} Insights, we obtained a lower $AL$ value compared to \textit{IM} Insights ($AL=0.161$ and $0.226$ respectively). To check the statistical significance of this difference, we followed the previous methodology and fit a GLMM model, but with the fixed effect being the experimental group instead of the specific interface. We confirmed that the $AL$ difference is significant with a pairwise comparison $p$\mbox{-value}$=0.004$). We also noticed from Figure~\ref{fig:pvaluesTable} that the best Far-reaching insight, \textit{FacesWithContext} ($AL=0.081$), performed more than twice better than the best Immediate Insight, \textit{Text} insight ($AL=0.206$) (ignoring the insights where the difference is not statistically significant). 
	Overall, these results demonstrate the superiority of our novel approach of \textit{FR} Insights. Nevertheless, \textit{IM} Insights are still significantly better than the \textit{BL} and \textit{DP} models. This goes in line with the findings of~\cite{Harbach:2014}, which showed the goodness of an approach similar to Immediate Insights in the case of Android permissions, even though they didn't have Delta Permissions as a building block.

	\subsection{Limitations}
	
	First, our design of the experiment abstracted several other factors that users take into account when installing apps. The interplay between ratings, app's brand, and permissions has been studied before (\cite{Kelley:2013} and~\cite{Harbach:2014}), and it might be worth revisiting in a future work in the context of our new permission models. Second, our experiment's advertisement included a mention of privacy as we wanted participants to focus on the app permissions. Evidently, this might have made participants more alert towards this issue. Both these points can imply that the real values of the $AL$ might be different in reality, where privacy might not be the main factor. 
	Nevertheless, even if the absolute values of $AL$ have been impacted, the relative advantages of new permission models still hold. 
	Moreover, we also note that the users in the $FR$ and $IM$ groups had to do more tasks than the $BL$ and $DP$ groups, which might have resulted in more user fatigue and habituation in the $FR$ and $IM$ groups. This was counteracted first via task randomization at design time and second by the very nature of insights that change at every step. For further validation, we computed the $AL$ values of Figure~\ref{fig:visuals_comparison}, considering only the first 5 tasks each user performed. We did not see any major deviation from the results with all tasks included. 
	Finally, our user recruitment strategy was primarily targeted towards our university's network, and our study was only for English speakers. It would be interesting to see how the results compare in a more general sample (linguistically, demographically and geographically).

	\section{PrivySeal: A Privacy-Focused App Store}
	
	\label{sec:privysealStore}
\begin{figure}[t]
		\centering
		\includegraphics[width=0.7\linewidth]{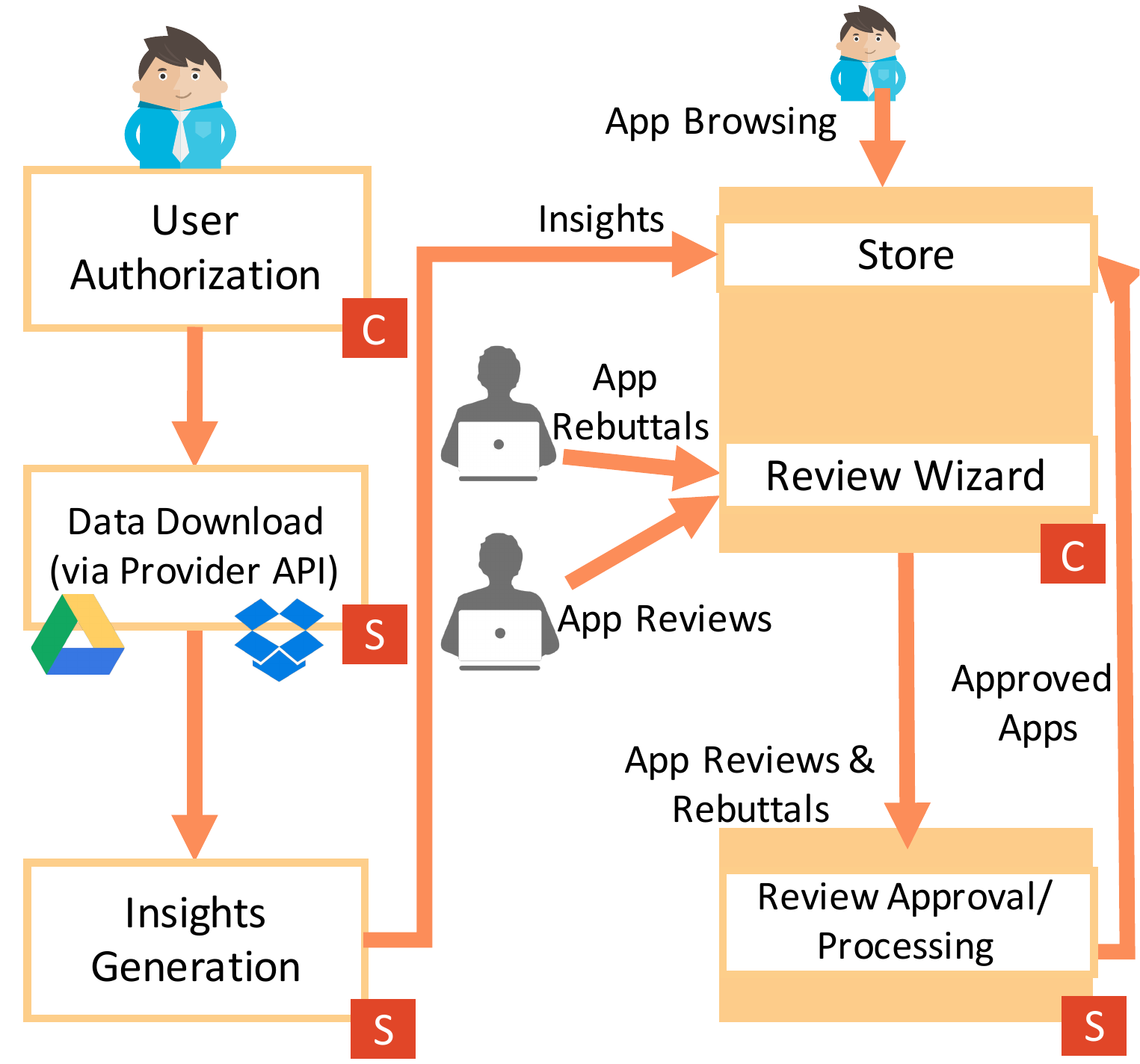}
		\caption{Component diagram of \privyseal (components labeled by $S$ are server-side and by $C$ are client-side)}
		\label{fig:architecture}
		
	\end{figure}

	Driven by the magnitude of the risk posed by over-privileged apps in Google Drive, we were motivated to bring the advantages of the Far-reaching Insights interface to the user community of this platform. 
	One approach towards achieving that would be for Google itself to implement a scheme similar to ours and to integrate it within the app authorization process. However, we decided not to wait and chose an alternative approach, which is independent of the company's plans and is ready for user utilization immediately. We built \privyseal, a privacy-focused store for Google Drive apps, which is readily available at \mbox{\url{https://privyseal.epfl.ch}}.
	\privyseal allows users to navigate a list of apps, click on those of interest, and check whether they are over-privileged via our \textit{FR} Insights interface. Users can also search by keyword for apps, specifying criteria such as the app being least-privileged.
	The component diagram for \privyseal is shown in Figure~\ref{fig:architecture}. 
	Similar to the APRs we conducted, we have included a ``Review Wizard'' inside \privyseal for indicating the requested, needed and unneeded permissions along with the alternative permissions the  developer could have used. This responsibility is currently given to a small set of expert developers and is moderated by the store administrators. Developers who would like to object to existing APRs of their apps can submit rebuttals.
	Currently, \privyseal has 100 apps and more than 1440 registered users, with a geometric mean of around 50 new users per month (whose vast majority is signing up out of interest in the app after reading article(s) about it).
	We finally note that \privyseal gets access, as is the case with other apps, to users' data to generate insights. Hence, users are assumed to trust the provider of such a ``Privacy-as-a-Service'' solution. However, this assumption of trust will hold if a solution such as \privyseal is hosted by the CSP itself (which already possesses the data), or an enterprise protecting its documents from 3rd party apps. The assumption of trust is also valid if the users choose to trust a \textit{single} entity (such as \privyseal) to protect themselves from \textit{multiple} other unaccountable over-privileged entities that they would otherwise be forced to trust.
	
	\section{Anatomizing Developers' Behavior}
	\label{sec:developers}
	
	After studying the users' privacy decisions, we now move to investigate the developers' landscape, building on the apps data that our store's users have installed. In total, we obtained data from 1440 registered users of our privacy store.

	\subsection{Current Developer Behavior}

	\label{sec:anatomDevCurrent}
	The ``\textsc{drive\_apps\_readonly}'' permission requested by our app allowed us to get the list of apps previously installed by users, along with the information that Google Drive API gives about the apps\footnote{For details, we refer the reader to: \url{https://developers.google.com/drive/v2/reference/apps}}. We found 662 unique apps installed by users in our dataset. For each app, we obtained the following:
	
	\textbf{i.} \textit{Access Level}:
	which indicates whether the app had \textit{Partial Access} or \textit{Full Access} to the user's drive \textit{on authorization time}. Since an app can change the permissions it requests from future users, our dataset had instances of the same app installed with different access levels by different users. We denote such apps as having an access level of \textit{Both}.
	
	\textbf{ii. } \textit{App Location}:
	which indicates whether the app is (1) on \textit{Google Chrome Web Store}, (2) on Google's \textit{Other Web Stores} (namely the Add-ons Stores and the Google Apps Marketplace for enterprises), or (3) \textit{Outside Web Stores} of Google. 
	This categorization is inferred by following the \textit{productUrl} field present in the app information, which either leads to one of the stores or is absent.
	
	Figure~\ref{fig:effectOfStore} shows how the apps in our dataset were distributed over the different locations and the number of apps requesting the different access levels.
	From this figure, one can observe the following:

	\begin{figure}[t!] 
		\centering
		\includegraphics[width=0.7\linewidth]{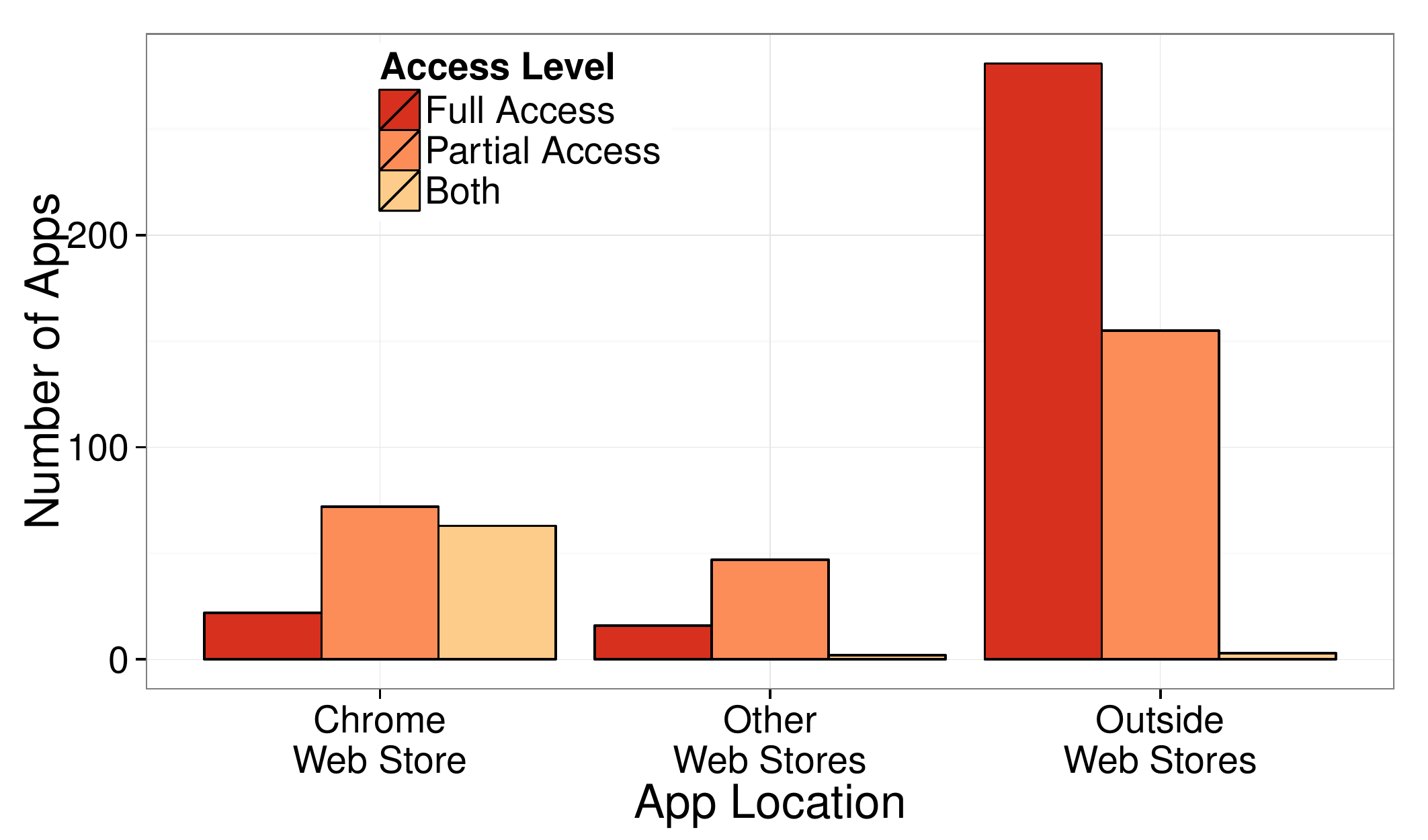}
		\caption{Change of access level with app location}
		\label{fig:effectOfStore}
	\end{figure}

	\tparagraph{Developers Changing Behavior}
	The first surprising outcome from this dataset is that around 40\% of apps on Chrome Web Store (63 apps) had \textit{Both} as access level, signifying that a lot of developers have changed the requested permissions at least once. In order to check the current access levels of these apps, we reviewed them one-by-one. We discovered that 59 of these apps (i.e., 94\%) have changed from requesting \textit{Partial Access} in the past, to requesting \textit{Full Access} currently. Hence, we can deduce that when developers change the access level, \textit{there is a high probability that it is associated with getting more data instead of the other way round}. Highlighting this change of access level on installation time can further serve for more informed user decisions.
	
	\tparagraph{Developer Deterrence through Official Stores}
	Apps outside the Web Stores requested \textit{Full Access} almost twice as much as they requested \textit{Partial Access} (281 full vs. 155 partial). This was not the case in the Chrome Web Store, where we observe only a slightly higher number of apps with \textit{Full Access} (81  full vs. 76 partial - counting apps that fall under the \textit{Both} access level but which currently request \textit{Full Access}). 
	So we can see that developers with apps outside the Web Stores are more prone to asking for \textit{Full Access}. This can be explained by the conjecture that the store acts as a medium where the apps receive more exposure. Hence, \textit{developers there are likely to be under the pressure of being evaluated through reviews and ratings, and thus tend to avoid abusing the permissions}, while developers outside Web Stores are under no such pressure. Although \textit{Full Access} does not necessarily mean that apps are over-privileged, our APRs have actually shown that 84\% of apps that request \textit{Full Access} are over-privileged apps.
	In the case of the Other Web Stores, the number of apps that requested \textit{Partial Access} is around thrice the number that requested \textit{Full Access} (16 full vs. 47 partial). This is mainly due to the fact that these Add-ons apps are generally expected to provide functionality for Google Docs (or other native Google file types), so deviating from this and requesting permissions for all Google Drive files will be easily detected by the community. Similarly, the community of enterprises, which is highly sensitive towards privacy will deter developers from requesting \textit{Full Access} in the Google Apps Marketplace.
	
	\tparagraph{Deterring Developers in the Wild}
	The majority of apps in our sample do not actually come from any Google Web Store (24\% from Google Chrome Store, 10\% from the Other Web Stores, and 66\% from outside the Web Stores). This is also the case for 75\% of the apps requesting full data access. These can be apps on other platforms, such as mobile platforms, for example, where there are other types of application stores. These apps can also be ones that are not present in any store but still have Google Drive integration. Hence, we can infer that improving the Chrome Web Store privacy indicators might not be a sufficient solution for deterring the majority of developers. \textit{There is a need for alternative solutions, focused on Google Drive permissions in specific, and independent of the various stores.}

	\subsection{Potential Developer Misbehavior}
	
	Although it is clear that full access to users' data can expose various far-reaching insights about the user, it is not completely apparent what seemingly benign permissions, such as metadata-only access can reveal about the user.  
	In the previous section, we have shown that the \textit{TopCollaborators} insight, which can be extracted just from the metadata, has resulted in an Acceptance Likelihood of $0.13$, which is around three times lower than what the current Google permission scheme (Baseline) attains. 
	Hence, users are remarkably deterred by seeing what they expose when they give access to their file metadata. Therefore, it is worthwhile to explore the potential of information leakage through metadata-only access.
	In this section, we show that metadata-only access on its own can allow developers to gather deeper insights about the user's topics and concepts of interest. This calls for extending the \textit{FR} Insights with such information in order to better inform the user about the potential risk of giving unneeded access to metadata. 
	Towards that, we analyze and compare insights inferred from users' file metadata to what can be inferred from file contents (the data). 
	
	Upon user \textit{u} signing up to our app, the following operations are executed as part of the data analysis:
	
	\indent \textbf{i. }For each analyzed file, the filename is processed by removing its extension and replacing punctuation marks by spaces.\\
	\indent \textbf{iii. }The names of all the analyzed files are grouped into a comma-separated list $L_{FN}(u)$\\
	\indent \textbf{iii. }Topic and concept analysis are applied to $L_{FN}(u)$. The service used for text analysis allowed us to extract topics in the form ``$a_1/a_2/ \ldots /a_n$'', representing a hierarchy among the labels (e.g., ``/law, government and politics/espionage and intelligence/surveillance'' or ``/finance/investing/venture capital'').
	In this section, we differentiate between \textit{General Topics} where we only consider $a_1$, and \textit{Specific Topics}, where we consider $a_n$. 
	Accordingly, \textit{General Topics} would indicate user's interest in \textit{law, government and politics} or \textit{finance} for example while \textit{Specific Topics} could indicate the user's interest in \textit{surveillance} or \textit{venture capital}.
	At the end of this step, we filter the results to restrict our analysis of metadata to a maximum of 3 \textit{General Topics}, 3 \textit{Specific Topics}, and 5 \textit{Concepts} for each user's list $L_{FN}(u)$.\\
	\indent\textbf{iv. }From the user's files' contents, we extract the top 5 \textit{General Topics}, top 10 \textit{Specific Topics}, and top 20 \textit{Concepts}. These choices are motivated by the general observation that one's \textit{Concepts} of interest are usually more in number than the \textit{Specific} abstract topics one cares about, which are in turn more than the \textit{General Topics} of interest. 
	
	For each user $u$, we compared the list $D(u)$ of labels (i.e., concepts/topics) extracted from the files' contents with the list $M(u)$ of labels extracted from the list $L_{FN}(u)$ of filenames.
	We selected precision as the evaluation metric as we are mainly interested in determining whether labels extracted from metadata serve as a good approximation of labels extracted from the data. 
	Inspired by the multi-label classification literature~\cite{Multilabel:2010}, we computed precision using the micro-averaging method, i.e., directly across all labels.
	A label occurrence is considered as true positive if it belongs to $M(u) \cap D(u)$ and a false positive if it belongs to $M(u) \setminus D(u)$. $tp(l)$ is the number of true positives for a label $l$, and $fp(l)$ is that of false positives, both across all users. Let $LT$ also be the set of all labels found across user's data and metadata. The overall precision is thus given by the following equation:
	\begin{equation}
		P_{micro}=\frac{\sum_{l \in LT}^{N} tp(l)} {\sum_{l \in LT}^{N} (tp(l)+fp(l))}
	\end{equation}
	We used this method instead of macro-averaging (i.e., computing the precision per label and then taking the average) because we are interested in estimating the users' interests more than the ability to predict each and every label.
	For this experiment, we only considered who signed in to our app and had at least 10 textual files with associated concepts/topics. Hence, our sample contained 200 users.
	Interestingly, the results for \textit{General Topics} indicate that 0.78 of the metadata labels across users match with their top 5 topics of interest.
	In the case of \textit{Specific Topics}, on average, nearly two of the three extracted metadata labels also appear in the top 10 \textit{Specific Topics} extracted from data ($P_{micro}=0.61$).
	Finally, the fraction of metadata \textit{Concepts} that also appear in the data is around one-third ($P_{micro}=0.31$).
	However, this does not necessarily imply that the other two-thirds of concepts appearing in the metadata are not relevant to the user. In fact, we have noticed that a lot of these metadata labels are semantically similar to those in the data. 
	
	In sum, we have observed that metadata on its own can be considerably accurate in revealing part of users' interests. 
	It can be easily abused by sophisticated adversaries who conceal their misbehavior through only requesting seemingly benign permissions (for metadata access). Therefore, this calls for extending the \textit{FR} insights in the case of metadata-only access to match the developer's potential. For instance, \textit{SharedInterests}, which was shown earlier to convey inferences from content, can also be used as an insight based on the collaborators and on the potential mutual topics inferred exclusively from the files' metadata.

	\section{Recommended Best Practices}
	\label{sec:privacyProtection}
	
	In addition to \privyseal, there are several steps that can serve to mitigate the potential of misbehavior in Google Drive and similar services. These solutions serve to help the user both before and after installing the apps.
	
	\tparagraph{Fine-Grained APIs}
	The availability of finer grained permissions (such as access for a specific file type) evidently reduces the amount of data in the hands of the developer and is in line with the principle of least privilege. One disadvantage of such detailed permissions is that they become more difficult for users to comprehend in a short amount of time. 
	However, providing developers with the means to request such fine-grained controls should not necessarily result in a more complicated interface. This can be achieved via multi-layered interfaces~\cite{schaub2015design}. For example, instead of the app indicating that it needs to ``View the files on your Google Drive'', it can indicate that it needs to ``View files of specific types in your Google Drive''. Users that are interested in knowing these file types can then click on an additional button (such as the info button \circled{i} in the current interface of Figure~\ref{fig:baseline_pdf_converter}).

	\tparagraph{Transparency Dashboard}
	A post-installation technique which can potentially deter developers from actually abusing the users' data is for the cloud platform to provide what we call ``Transparency Dashboards''. These dashboards allow the user to see which files have been downloaded by each 3rd party app and when such operations took place. Such a monitoring solution for all apps can only be achieved by the platform itself. 
	
	\tparagraph{Insights Based on Used Data}
	Unlike external solutions (e.g., ours) that can only determine what data can be \textit{potentially} accessed, the cloud platform can provide users with insights based on the data that developers have \textit{previously} downloaded. Such an interface will help users better pinpoint adversarial apps that needlessly retrieve files outside the scope of their functionality.
	
	\tparagraph{A Privacy Preserving API Layer}
	It is not uncommon nowadays to find APIs that work as an additional layer on top of one or more existing cloud APIs (e.g., Cloud Elements Documents Hub).
	Hence, one solution to build a privacy-preserving API is to create it as a layer on top of one or multiple existing platforms' APIs. This new API can provide finer grained access control, allow permissions reviews from the community, and implement transparency dashboards. By building this layer on top of existing cloud APIs that already offer various services, one can circumvent the problem of attracting developers who might otherwise be loathe to using a solution that only serves to protect privacy.

	\section{Related Work}
	
	To our knowledge, this is the first work that studies the problem of user privacy in the context of 3rd party apps on top of cloud storage providers. Other works have previously studied the problem of direct information sharing to providers themselves (e.g.,~\cite{harkous2014c3p}).

	\subsection{Privacy in Other App Ecosystems}
	
	In the case of other ecosystems, there are related works that have studied the current state of privacy notices (e.g. ~\cite{Chia:2012, Huber:2013, Felt:2011, Pandita:2013}).
	For instance, Chia et al., conducted a large-scale analysis of Facebook apps, Chrome extensions, and Android apps to study the effectiveness of user-consent permissions systems~\cite{Chia:2012}. They observed that the community ratings are not reliable indicators of app privacy in these ecosystems and showed evidence of attempts at misleading users into granting permissions via free apps or apps with mature content.
	Huber et al., developed \textit{AppInspect}, a framework for automating the detection of malpractices in 3rd party apps within Facebook's ecosystem and used network traffic analysis to spot web trackers and identify leaks of sensitive information to other third parties~\cite{Huber:2013}.
	The case of 3rd party apps in Google Drive differs from these platforms in that it is not possible to perform large-scale analysis, firstly due to the absence of a standard application format and secondly due to the difficulty of automatically finding the triggering button for permission requests in different apps. 
	Aside from the above, client-side traffic analysis is not sufficient to detect all cloud data leaks as the apps can send data to third parties after it arrives at the server side, to which outsiders do not have access.

	\subsection{Improving Current Privacy Notices}
	
	A few works have recently suggested improvements to the existing permissions schemes, with a special focus on the case of Google Play Store. 
	Kelly et al., argued that the privacy information should be a part of the app decision-making process and should not be left till after the user makes her decision~\cite{Kelley:2013}. Hence, they appended a list of ``Privacy Facts'' to the  app description screen, textually indicating that the app, for example, collects contacts, location, photos, credit card details, etc., and found that it assisted users in choosing apps that request fewer permissions. 
	Harbach et al., proposed to integrate examples from user's data in the permissions request screen to expose the data apps can get access to~\cite{Harbach:2014}. This involved showing random pictures, call logs, location, and contacts from user's data that correspond to each permission. 
	Another related work in the context of Facebook is that by Wang et al.~\cite{Wang2013}, who introduced the ``Privacy Nudges'' technique to aid users while posting statuses to Facebook through showing random profile pictures of friends who can see the post, introducing a time interval before the actual post is sent, or showing the post sentiment. 
	In this work, we go further, and we show that well-crafted visuals showing far-reaching insights extracted from users' data can be more effective than randomly selected data. We also show through pairwise comparisons among the insights themselves that the choice of the displayed insight highly affects the interface's effectiveness.
	It is also worth mentioning that, in our experiments the number of users who were involved with their personal accounts in the experiment was more than five times the number of users in~\cite{Harbach:2014} and~\cite{Kelley:2013}. Furthermore, we also provide a readily available solution for the public in the form of a privacy-focused app store.

	Moreover, our work is in line with the best practices recommended by the recent work of Schaub et al., who developed a design space for privacy notices to assist researchers in increasing the impact of their schemes~\cite{schaub2015design}. For instance, we implemented the multi-layered notice concept by showing data of textual and visual modalities. We also developed various visuals to ensure that the permissions dialog is \textit{polymorphic}, which was also shown recently to have an effect on reducing the habituation effect in the user's brain~\cite{Anderson:2015}. 
	Personalizing warning notices, as we do in this paper, has been studied before in the context of LED signs~\cite{Wogalter1994233} and was shown to significantly increase compliance compared to impersonal signs.

	\section{Conclusion and Future Work}
	
	In this paper, we characterized the various factors that have an impact on user privacy in the ecosystem of 3rd party apps for the cloud. We considered Google Drive as an example case study and comprehensively anatomized the ecosystem from the viewpoint of users, developers, and the cloud provider. For users, we carefully devised a set of experiments and tested existing and novel risk communication models to analyze the factors that influence users' decisions in app installation. Our results provide interesting insights into how user privacy can be improved and how CSPs can develop better risk indicators. We also presented a privacy aware store for cloud apps, which already has over 1440 registered users. From our store users and people who took part in our experiments, we had the unique and unprecedented opportunity to first-hand study real users cloud data. Based on this data, we were able to characterize the current behavior of 3rd party app developers and also point out avenues for developer misbehavior. Finally, based on our analysis, we provided several suggestions for CSPs that can help in safeguarding users' privacy and protecting their data from needless leakage and exploitation. 
	In the future, we aim to build on \privyseal and develop a recommendation system that suggests apps of similar functionality but superior privacy. 
	We are also in the process of integrating Personalized Insights in the scenario of user-to-user sharing privacy in \textit{StackSync}, which is an open source cloud platform with a significant number of users. 
	Finally, it would be interesting to study how our findings on the best risk indicators generalize to other ecosystems, such as Android or iOS.

	\section*{Acknowledgments}
	We would like to thank the anonymous reviewers for their insightful comments. The research leading to these results has received funding from the EU in the context of the project \textit{CloudSpaces}: Open Service Platform for the Next Generation of Personal clouds (FP7-317555).

	\bibliography{IEEEabrv,references_to_compile}
	\bibliographystyle{myIEEEtranS}
	
	\appendix

	\section{Review Process and Data}
	
	\label{apndx:reviewData}
\begin{figure}[b!]
		
		\includegraphics[width=0.8\columnwidth]{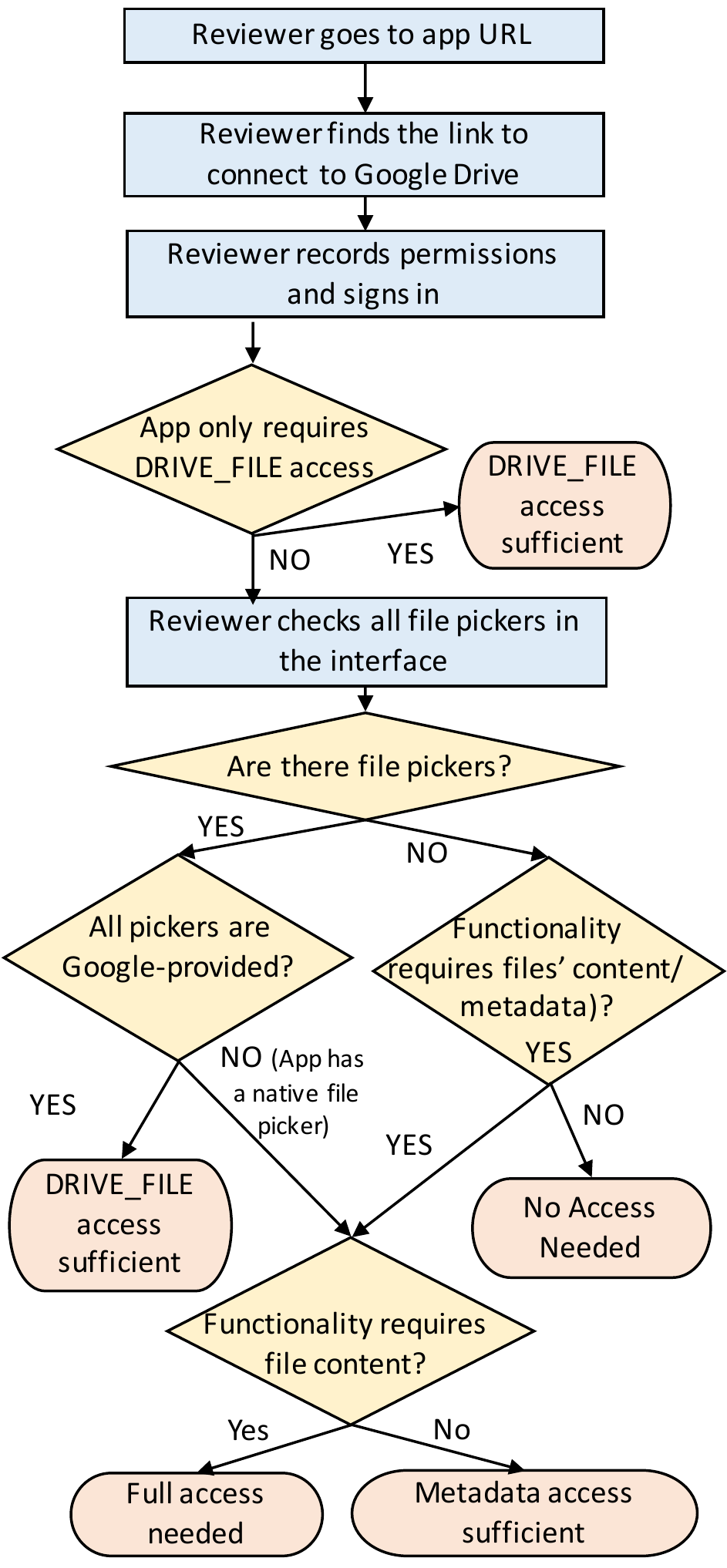}
		\caption{Flowchart of the APR process, inspired by Google Drive guide for choosing authentication scopes \\({\url{https://developers.google.com/drive/v3/web/about-auth}})}
		\label{fig:review}
	\end{figure}

\begin{figure*}[t!] 
		\minipage{0.25\linewidth} 
		\includegraphics[width=0.8\linewidth]{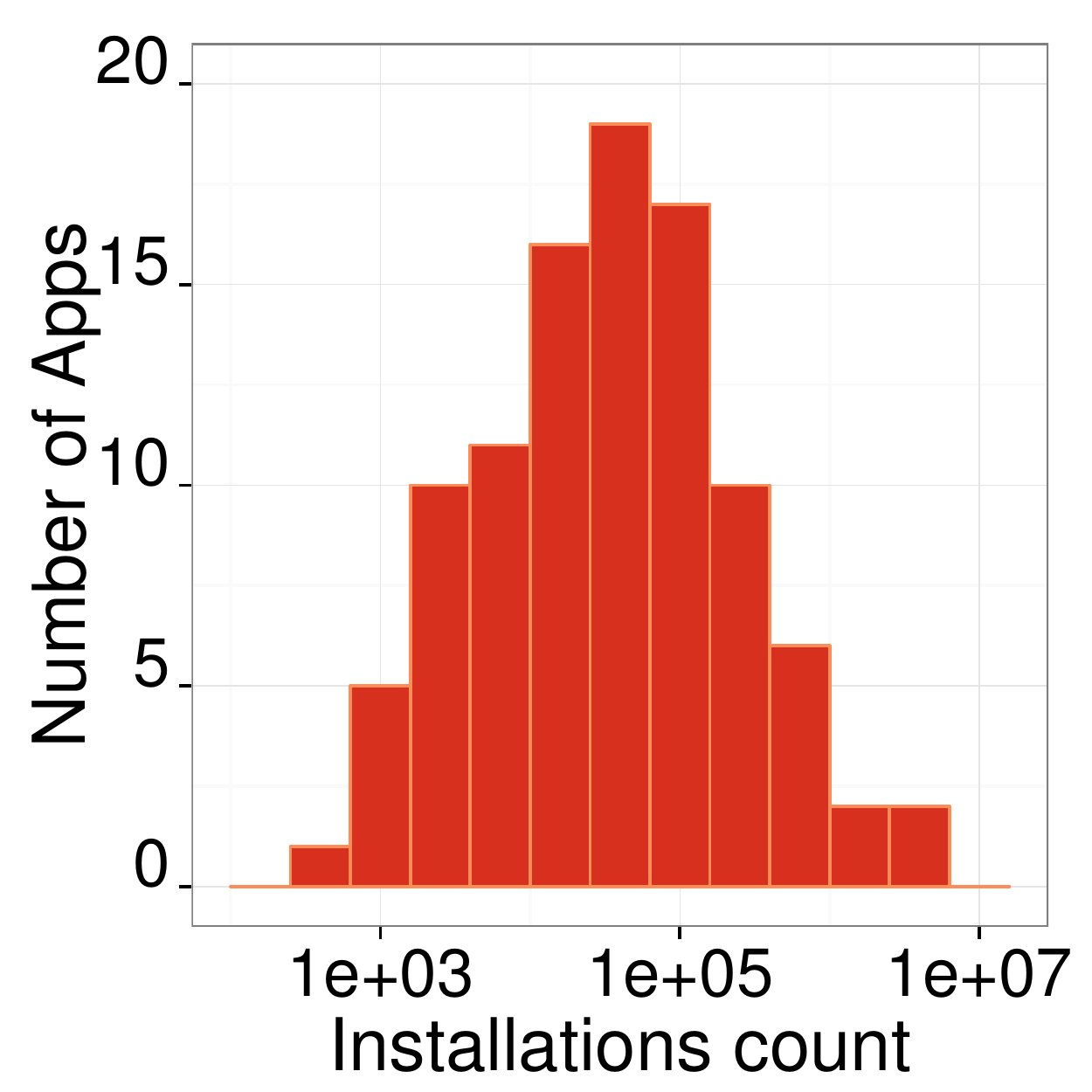}
		\caption{Distribution of the app installation counts (on a log scale) in the reviewed dataset}
		\label{fig:appUsersHist}
		\endminipage\hfill
		\minipage{0.25\linewidth} 
		\includegraphics[width=0.8\linewidth]{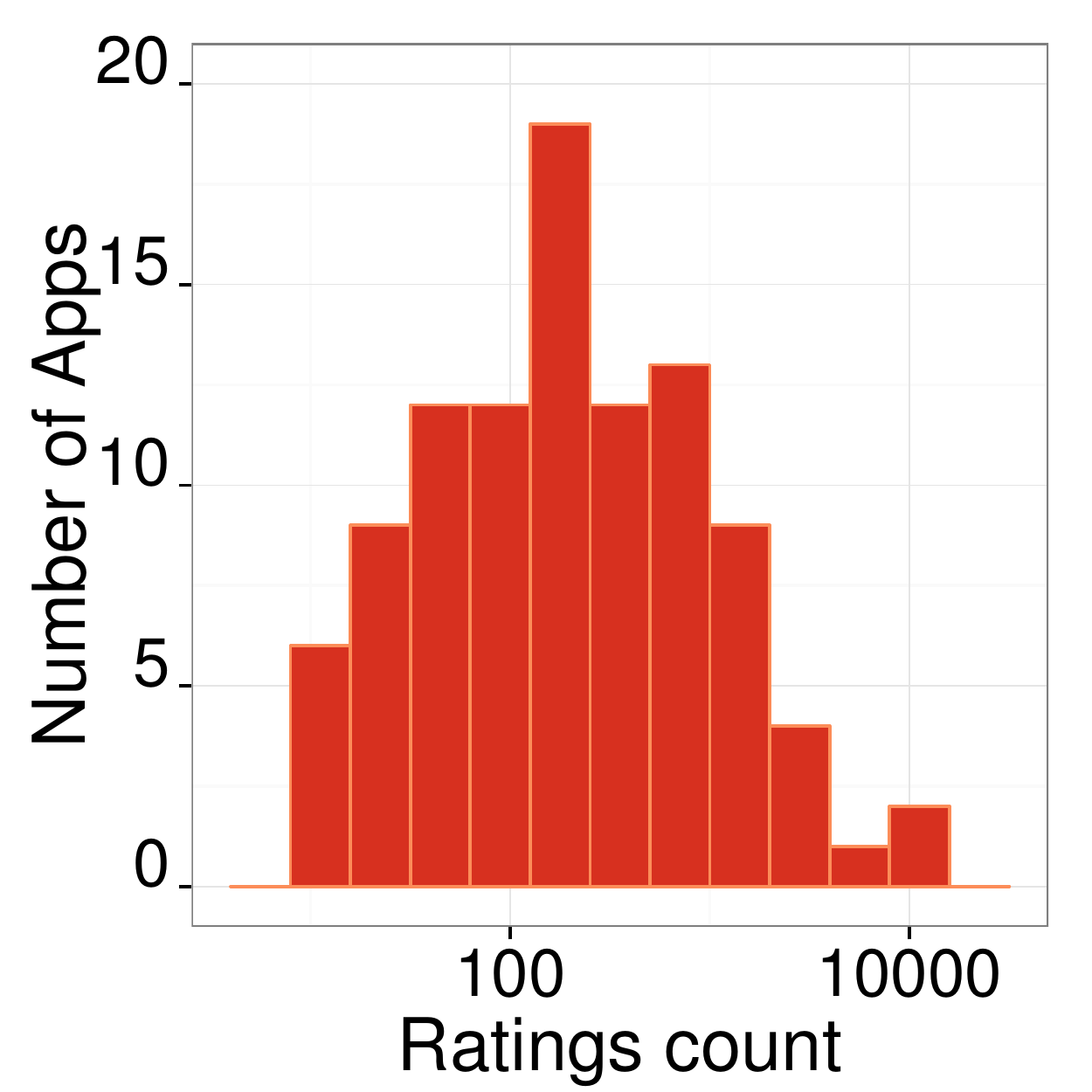}
		\caption{Distribution of the app rating counts (on a log scale) in the reviewed dataset}
		\label{fig:appRatingsCountHist}
		\endminipage\hfill
		\minipage{0.25\textwidth} 
		\includegraphics[width=0.8\linewidth]{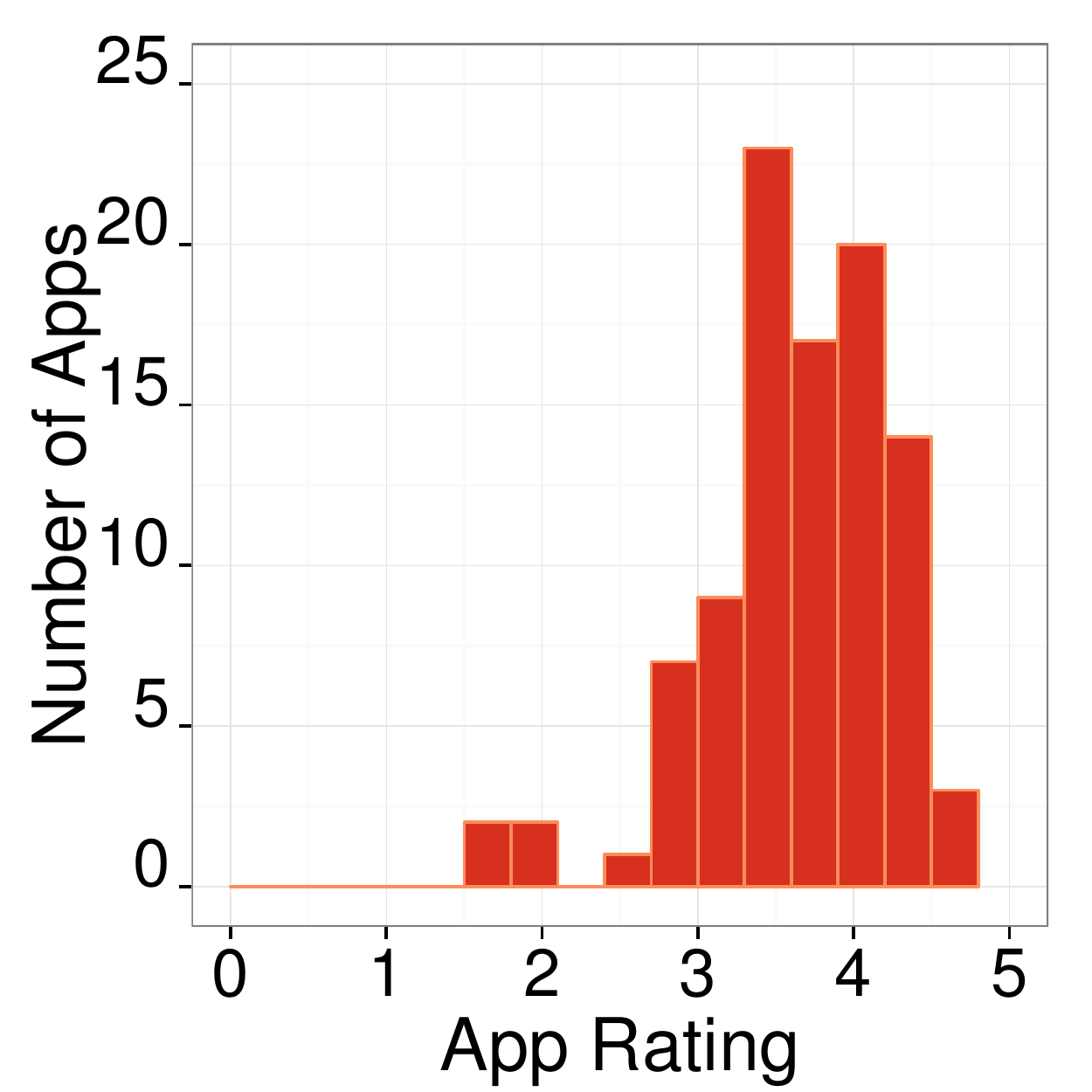}
		\caption{Distribution of the app rating values in the reviewed dataset}
		\label{fig:appRatingsValHist}
		\endminipage \hfill
		\minipage{0.25\textwidth} 
		\includegraphics[width=0.8\linewidth]{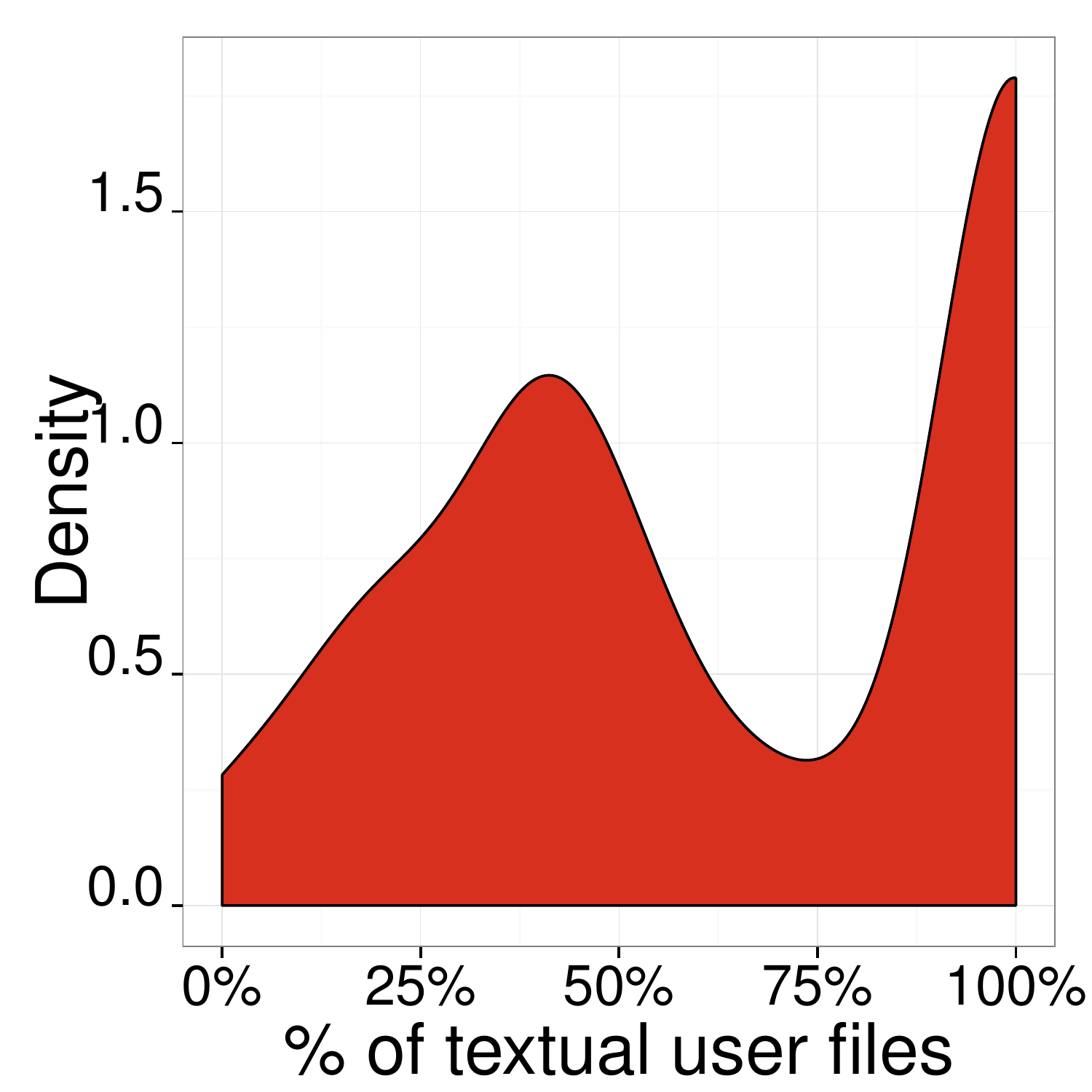}
		\caption{Density plot of the percentage of textual files for our experiment's users}
		\label{fig:userFileStats}
		\endminipage

	\end{figure*}

	Figure~\ref{fig:review} shows the flowchart of the APR process described in Section~\ref{sec:revProcess}. 
	Figure~\ref{fig:appUsersHist} shows the distribution of the installation counts of apps in our dataset on a log scale, where it is clear that the apps follow closely a normal distribution (this has been individually confirmed using q-q plots). The average number of installations was 194,600 and the median was 29,350. Figure~\ref{fig:appRatingsCountHist} shows that the number of ratings follows a similar distribution, with a mean of 736 and a median of 181. The ratings value distribution is shown in Figure~\ref{fig:appRatingsValHist}, with a mean of 3.66 and a median of 3.72. Overall, this shows the diversity of the apps in our APRs dataset and that represents a wide range of apps.

	\section{Detailed Insights Description}
	
	\label{apndx:Insights}
	
	In this appendix, we detail the different Far-reaching insights that were presented in Section~\ref{sec:farInsights} and explain the algorithms used for generating each of them. 
	Towards that goal, we highlight two file categories of interest: (1) textual files, such as PDF documents, word-processing documents, spreadsheets, presentations, text files, etc., and (2) image files, such as JPEG, PNG, TIFF, etc.
	We represent the set of textual files as $TF =TF_1,TF_2,..,TF_K$ and the set of image files as $IF_1, IF_2, ..., IF_L$.

	\tparagraph{Entities, Concepts, and Topics (ECT)}
	
	\textbf{i. Entities:}
	We get the \emph{top named entities} (e.g., people, places, companies, etc.) present in the user's textual files. Such entities are recognized using Named Entity Recognition (NER), which is a traditional problem in natural language processing that involves locating and classifying elements in text into pre-defined categories~\cite{Downey:2007}.
	For this task, we perform text extraction on each file, and we then pass the text to a AlchemyAPI's service.
	Given the text of file $TF_j$, this service returns a set of entities, along with the frequency of occurrence $f_{i,j}$ of each entity $e_{i}$ in $TF_j$. 
	We normalize this frequency for each entity by dividing it by $\mbox{\textit{fmax}}_j$, which is the frequency of the most recurrent entity in $TF_j$:
	\begin{equation}
		\mbox{\textit{fnorm}}_{i,j}=\frac{f_{i,j}}{\mbox{\textit{fmax}}_j}
	\end{equation}
	Then, we compute an overall score for entity $e_i$ across all the files in $TF$, by summing its individual normalized frequencies:
	\begin{equation}
		score(e_i)= \sum\limits_{j=1}^{K} \mbox{\textit{fnorm}}_{i,j}
	\end{equation}
	As shown in Figure~\ref{fig:entities}, we visualize the entities with the highest scores as a set of circles, each of a diameter proportional to the score of the corresponding entity. Different types of entities (e.g., people, places, etc.) have different circle color.

	\noindent\textbf{ii. Concepts:}
	We also extract concept tags from users' documents. These concepts are high-level abstractions, not necessarily mentioned in the text. 
	AlchemyAPI was again used for this task, returning, for each file $TF_j$, a 
	set of concepts, each denoted as $c_i$ along with a relevance score $r_{i,j} \in [ 0,1]$.
	We used the following scoring method to rank the concepts across the user's documents:
	\begin{equation}
		score(c_i)= \sum\limits_{j=1}^{K} r_{i,j}
	\end{equation}
	Similar to the case of entities, we represent concepts by circles, each of a diameter proportional to the score of the concept.
	
	\noindent\textbf{iii. Topics:}
	Topics are used to classify documents into high-level categories, such as technology, art, business, etc.
	We used AlchemyAPI, which returns a maximum of 3 topics per file $TF_j$ (each denoted as $t_i$), along with a relevance score $r_{i,j} \in [ 0,1]$ for each of them. A topic comes in the form of ``$a_1/a_2/ \ldots /a_n$'', representing a hierarchy among the labels (e.g., ``/hobbies and interests/astrology'' or ``/finance/investing/venture capital'').
	In order to extract the top topics based on a user's documents, we use the same scoring method as that of concepts:
	\begin{equation}
		score(t_i)= \sum\limits_{j=1}^{K} r_{i,j}
	\end{equation}
	We represent topics by circles, similar to the case of entities, where the diameter of a circle is proportional to the score of the topic. Topics sharing the top level label are colored similarly.

	\tparagraph{Sentiments}
	For each entity that occurs in $TF$, it is possible to also estimate whether the text relays a positive, neutral, or negative sentiment about that entity. 
	Towards that end, we use the sentiment analysis service of AlchemyAPI. For each $TF_i$, we select the entities labeled with positive or negative sentiments (each such entity also has a sentiment score $s_{i,j} \in [ -1,1]$ with 1 corresponding to the most positive sentiment and -1 to the most negative one.). 
	We then compute the overall sentiment score $s_{i}$ of entity $e_{i}$ across the all the user documents $TF$:
	
	\begin{equation}
		s_{i}= \sum\limits_{j=1}^{K} s_{i,j}
	\end{equation}
	The sentiments with the highest positive and negative scores are then shown to the user, as was presented in Figure~\ref{fig:sentiments}.

\begin{figure*}[t!] 
		\centering
		\minipage{1\linewidth} 
		\centering
		\includegraphics[width=0.65\linewidth]{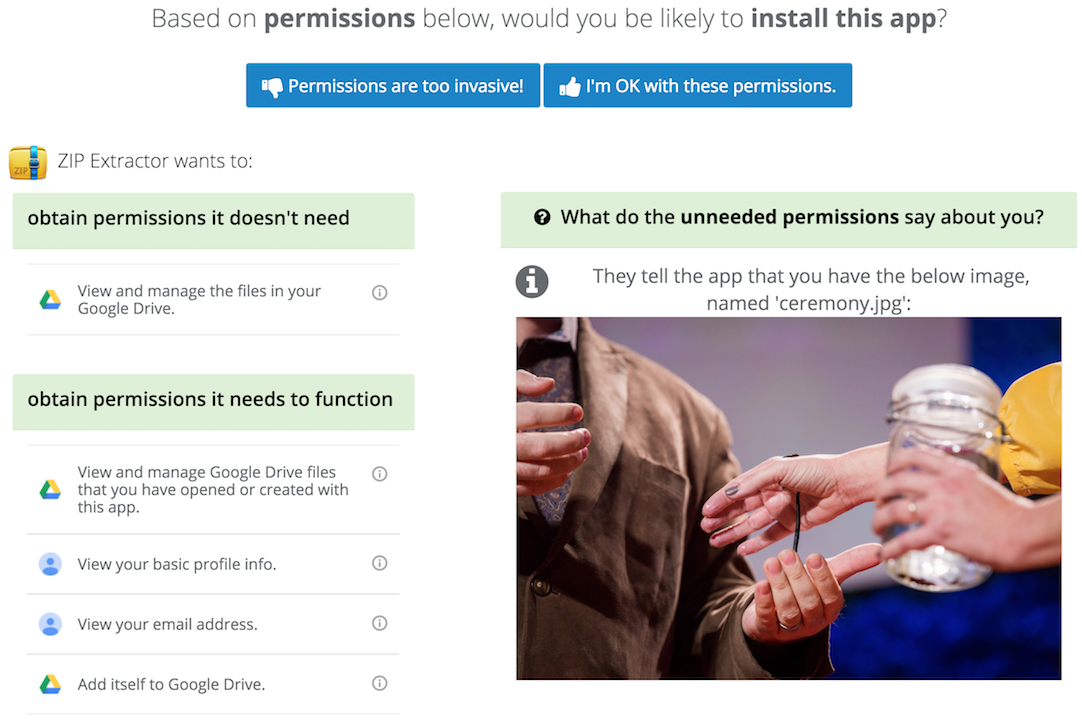}
		\caption{Example of the interface shown to users of the \textit{IM} group, with the decision dialog on top}
		\label{fig:immediate}
		\endminipage\\
		\minipage{1\linewidth} 
		\centering
		\includegraphics[width=0.65\linewidth]{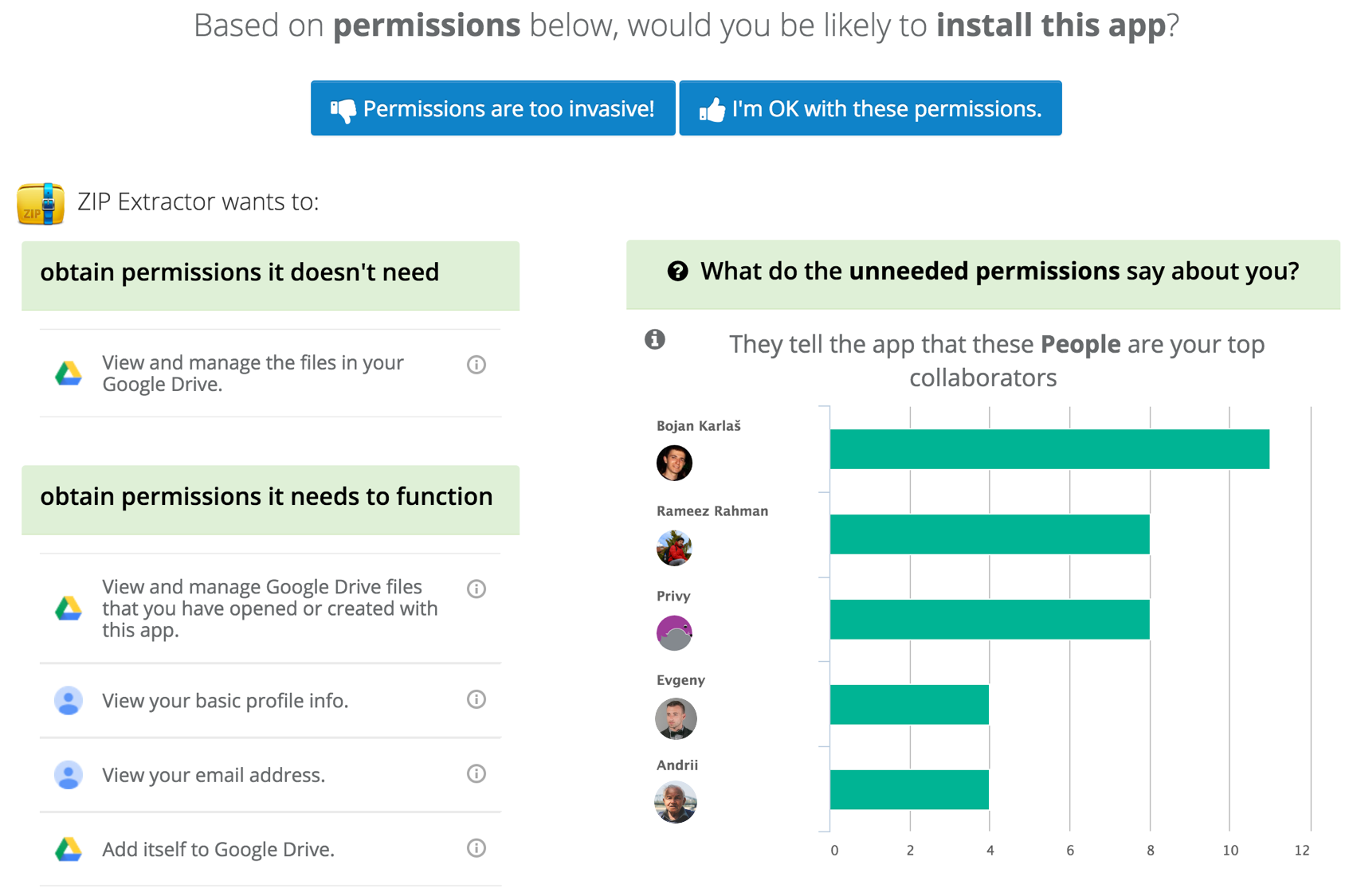}
		\caption{Example of the interface shown to users of the \textit{FR} group, with the decision dialog on top}
		\label{fig:far}
		\endminipage\\\minipage{1\linewidth} 
		\centering
		\includegraphics[width=0.8\linewidth]{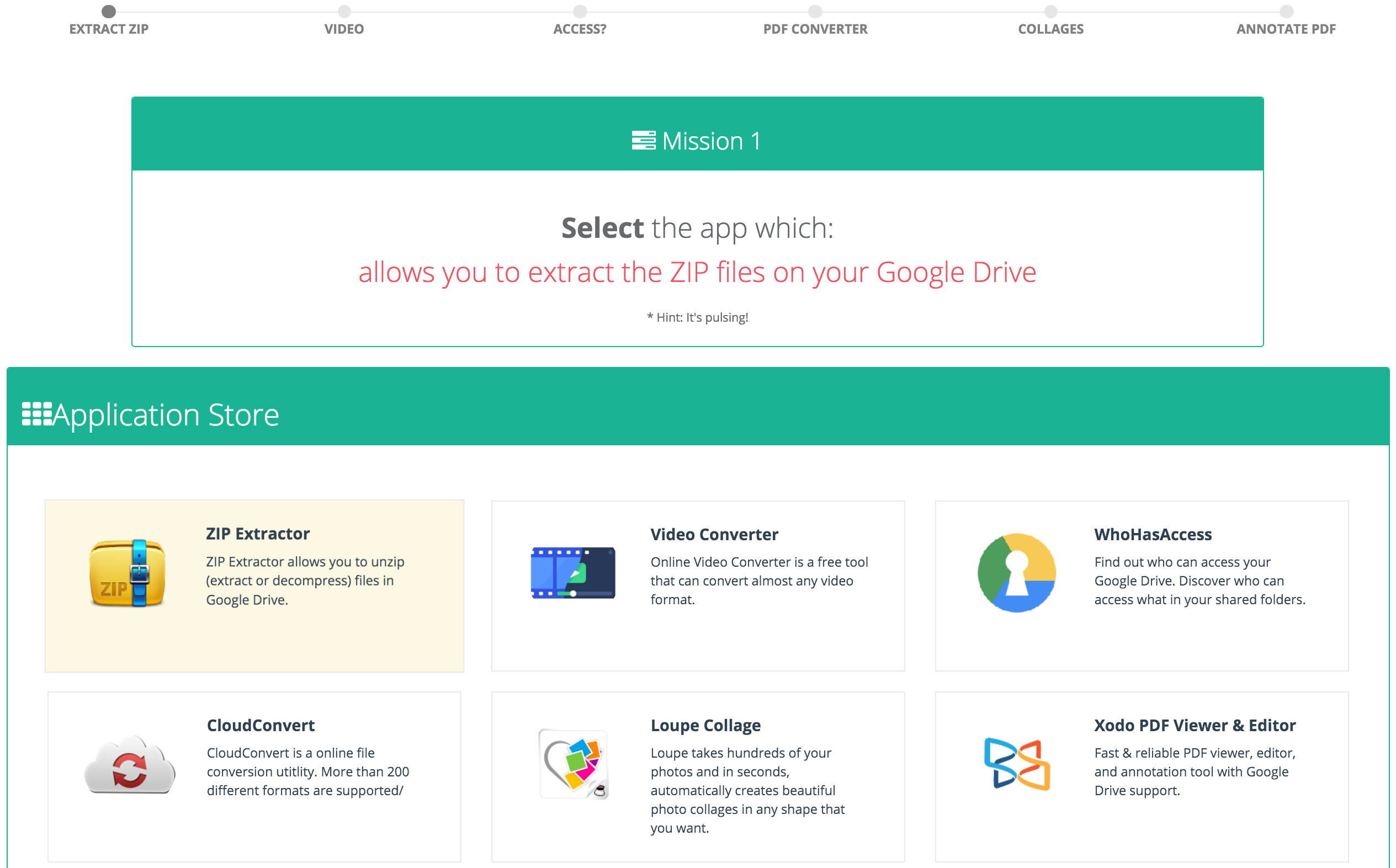}
		\caption{Task interface presented for the users in the experiment,\\ where they had to select the app satisfying the given purpose (already highlighted for them)}
		\label{fig:task}
		\endminipage
	\end{figure*}

	\tparagraph{Top Collaborators}
	We define collaborators as people who share files with the user, regardless of who initiates the sharing operation. 
	In the interface, this insight is visualized as a horizontal bar chart of the top collaborators with the bar lengths representing the relative frequency of the user's collaboration with each of them (an instance of this visual appears in Figure~\ref{fig:far} in the context of our experiment).
	
	\tparagraph{Shared Interests}
	In this insight, we try to represent the user's mutual topics of interests with a group of people. 
	Towards that end, we perform the following steps:
	\begin{itemize}
		\item
		We determine the top topics as we have done in the \textit{ECT} insight. 
		\item
		Then we select from these topics a subset $S_t$ that only includes the ones which appeared in shared files. 
		\item
		Via Google Drive API, we extract, for each topic $t_i$, a list $U(t_i)$ of collaborators (based on files it appeared in).
		\item
		We select from each $U(t_i)$ the most frequent collaborators (i.e., those appearing in most documents with this topic).
	\end{itemize}
	Users then get a visualization similar to Figure~\ref{fig:sharedInterests}, where we show the three top topics from $S_t$ along with the top collaborators for these topics.
	
	\tparagraph{Faces with Context}
	This insight shows a group of faces, representing the most frequent people appearing in the user's images, alongside the concepts that appear in the same images. 
	In order to achieve it, we performed two steps:
	
	\textbf{i.} Face clustering: It is evident that showing the user random faces detected in her photos will not create the same effect as when these faces are actually people she cares about. Our plan to achieve the latter case involves three steps:
	
	\begin{itemize}
		\item
		We use a face clustering algorithm in order to group together photos of the same person. As a result, we get a list of groups $G$, where each group $G_i \in G$ is comprised of the faces that belong to a person identified as $p_i$. The algorithm used is by Zhu et al.,~\cite{ChunhuiZhu:2011}  implemented by the OpenBR framework.~\cite{Klontz}

		\item
		From each group $G_i$, we exclude the faces with width (height) less than $\frac{1}{15}$ of the total image width (height).
		\item
		We exclude groups with less than 3 faces in total.
		\item 
		We sort the groups by the number of faces in each of them.
	\end{itemize}
	
	\vspace{-\baselineskip}
	\textbf{ii.} Image concept recognition: 
	In order to identify the concepts inside each photo, we used a classifier from the Caffe library~\cite{jia2014caffe}.
	The classifier uses a pre-built deep learning network, that is based on the architecture used by Krizhevsky et al.,~\cite{krizhevsky2012imagenet} that won the Imagenet 2012 contest.
	
	Based on the above, we show the user the top groups (i.e. with most faces) along with the most recurring concepts in these groups (as in Figure~\ref{fig:faceswithcontext}).
	
	\tparagraph{Faces on Map}
	
	This insight (shown in Figure~\ref{fig:facesonmap}), consists of showing the faces of people overlaid on a map, centered at the geographical area where these faces appeared. Below the map is a list of the top concepts that appeared in the photos taken in that area. In our actual implementation, the visual is animated, moving between different areas to show the user the places that different photos were taken at. In order to construct this visualization, we had to cluster the images into different geographical areas. For that, we used the OPTICS algorithm (Ordering Points to Identify the Clustering Structure) by Ankerst et al.,~\cite{Ankerst:1999}. OPTICS allows finding density-based clusters in spatial data and is tailored for detecting meaningful clusters with data of varying density. After getting the cluster results, the zoom level on the map is animated to show one cluster to the user at a time.

	\begin{figure}[t!] 
		\minipage{0.47\linewidth} 
		\includegraphics[width=\linewidth]{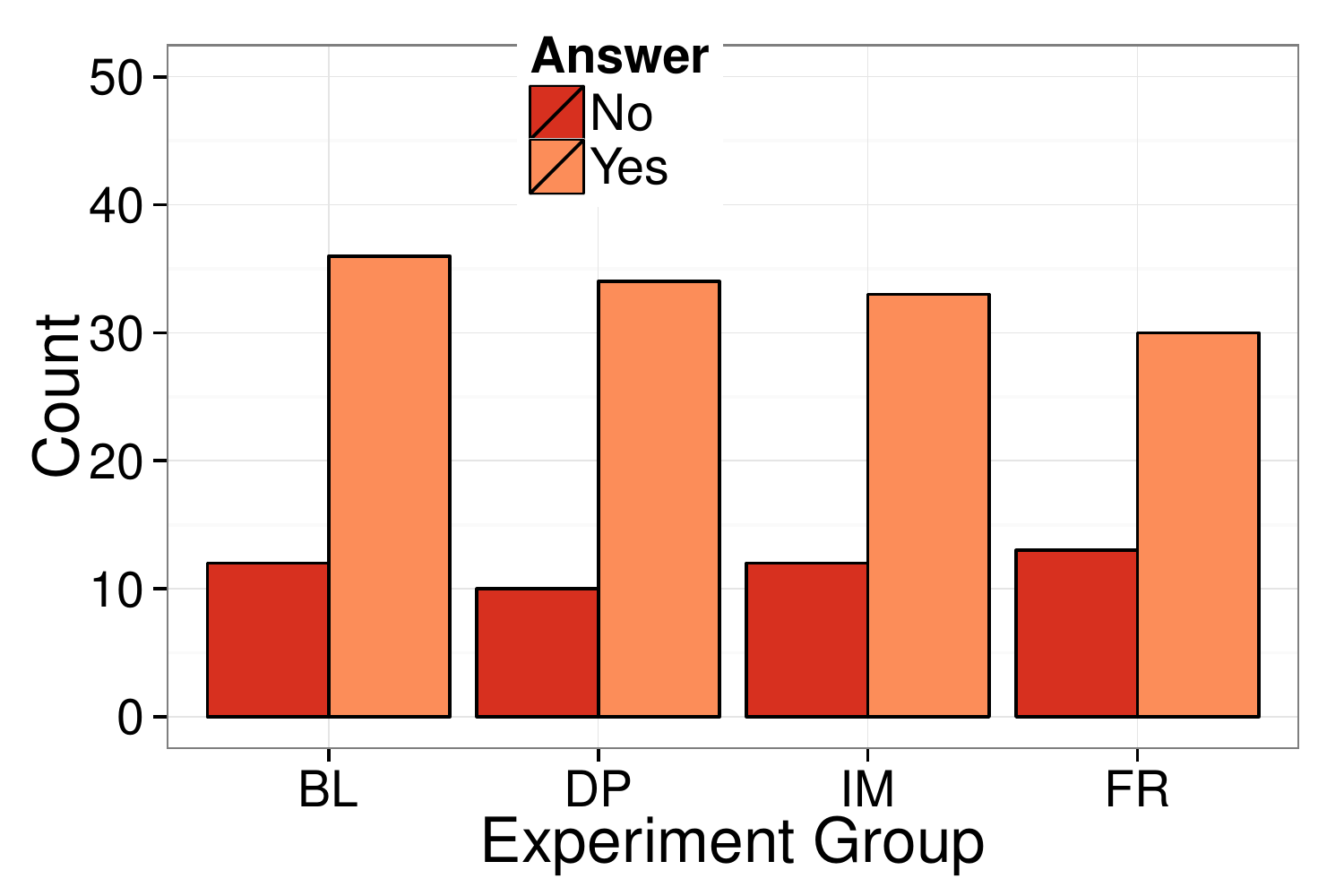}
		\caption{\textbf{Q:} I understand what the different Google permissions mean.}
		\label{fig:understanding}
		\endminipage\quad
		\minipage{0.47\linewidth} 
		\includegraphics[width=\linewidth]{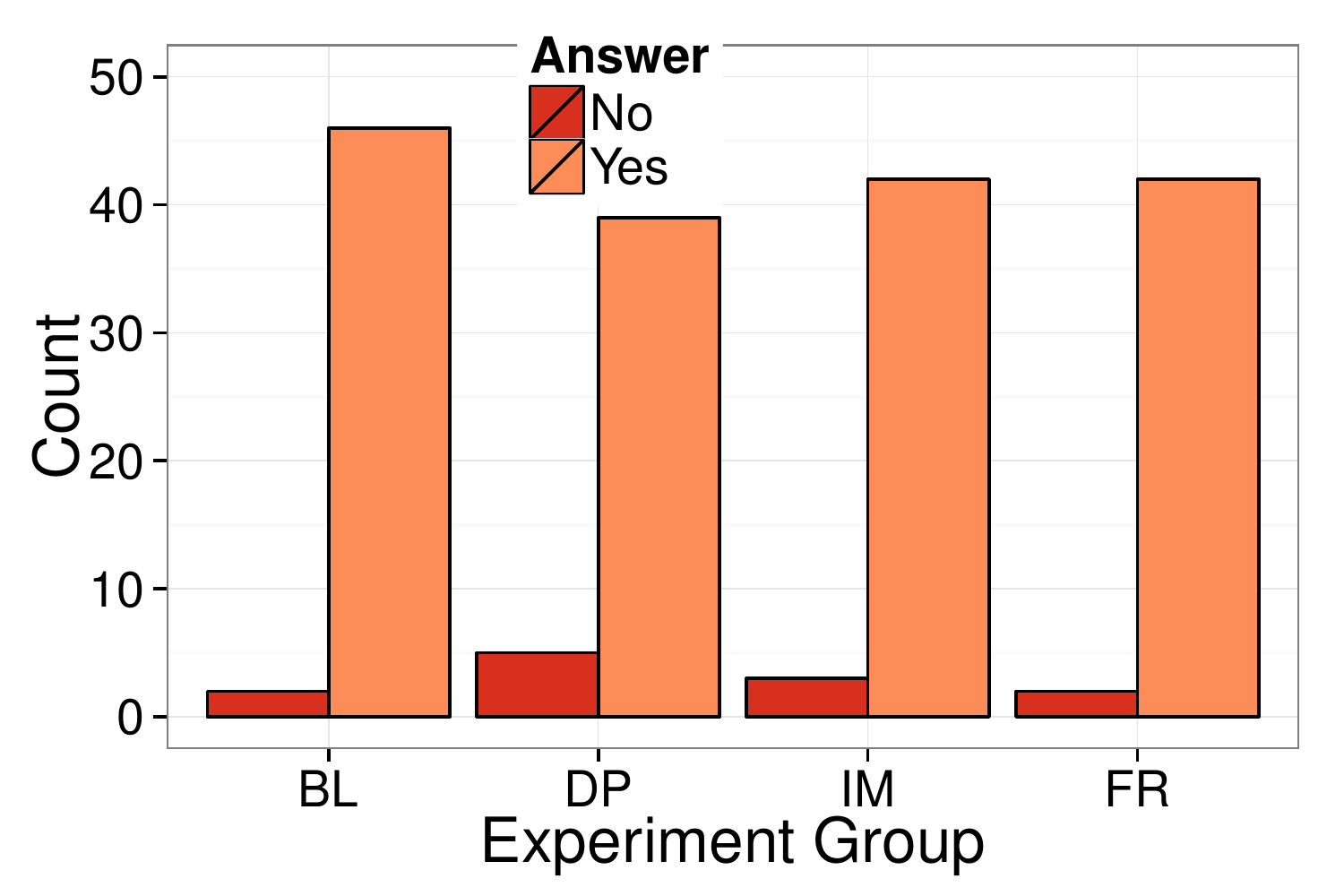}
		\caption{\textbf{Q:} I found the interface in these missions intuitive.}
		\label{fig:intuitive}
		\endminipage
	\end{figure}
	\begin{figure}[t!] 
		\minipage{0.47\linewidth} 
		\includegraphics[width=\linewidth]{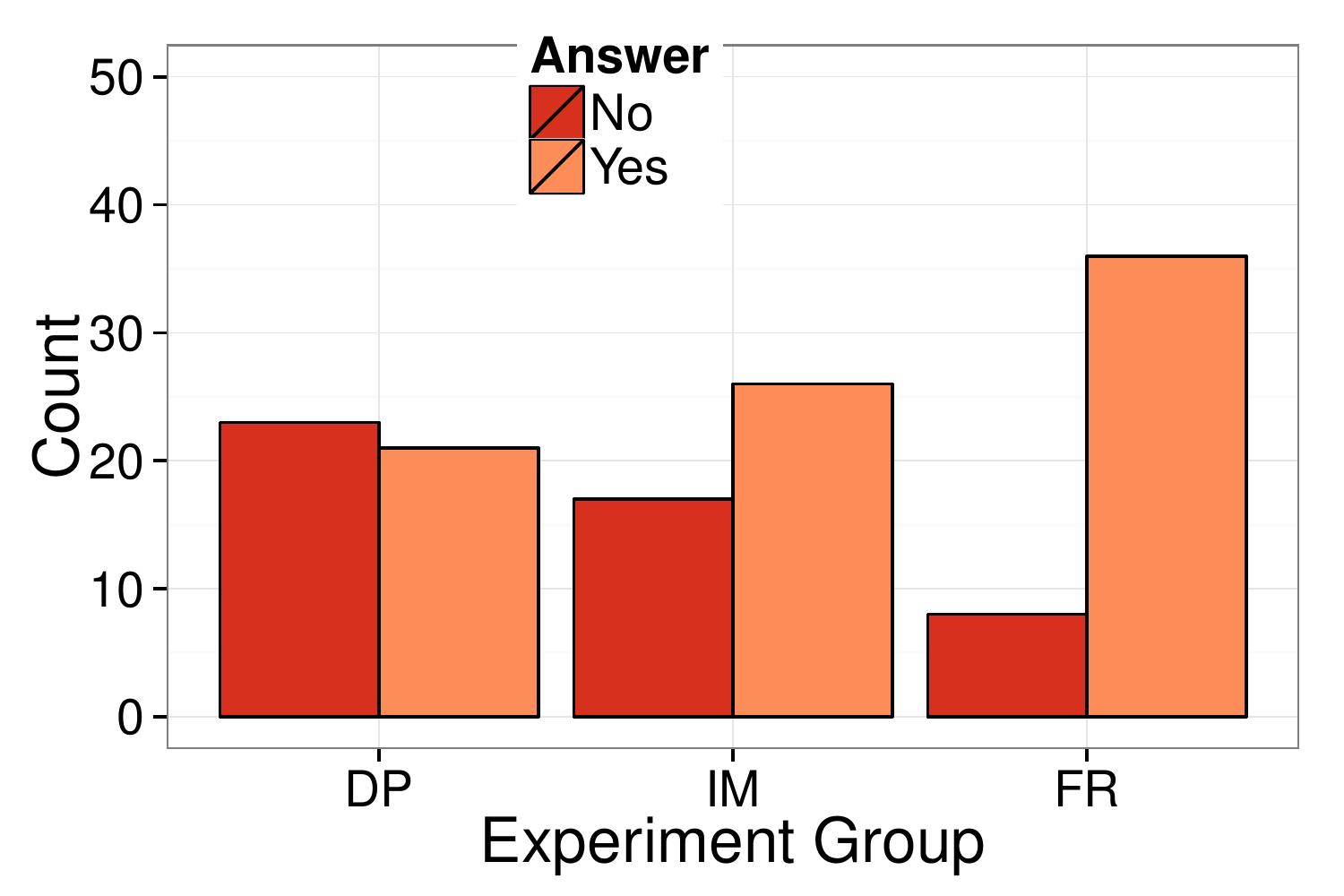}
		\caption{\textbf{Q:} I was surprised that apps know more about me than I expected.}
		\label{fig:surprised}
		\endminipage\quad
		\minipage{0.47\linewidth} 
		\includegraphics[width=\linewidth]{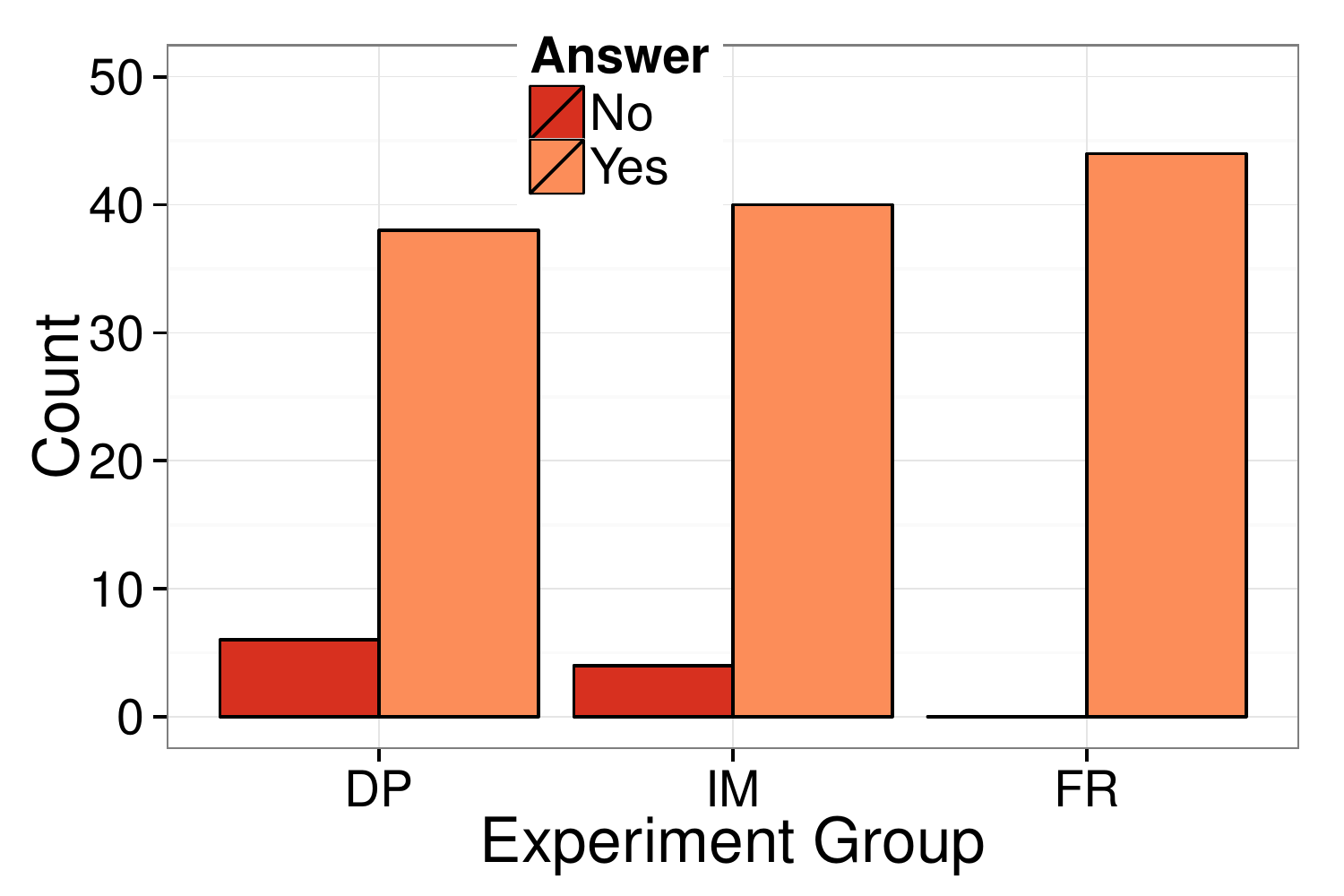}
		\caption{\textbf{Q:} I would be interested in such a Store before installing real apps.}
		\label{fig:realApps}
		\endminipage
	\end{figure}

	\vspace{-\baselineskip}
	\section{Further Experimental Details}
	\label{appndx:experiment}
	
	Figure~\ref{fig:userFileStats} shows the density plot for the percentage of textual files in the user's Google Drive (out of both image and textual files). Although there is a large fraction of users with no image files, there are many users with a balanced fraction of textual and image files. Figure~\ref{fig:task} shows the task interface given to users during the experiment, as explained in Section~\ref{sec:Exp1}. Users read the task goal and select the highlighted app. Then they are directed to a permissions interface corresponding to their group. Figures ~\ref{fig:immediate} and~\ref{fig:far} show screenshots of the interface for the $FR$ and $IM$ experimental groups. The cases of $BL$ and $DP$ groups are similar to Figures~\ref{fig:baseline_pdf_converter} and~\ref{fig:delta_example}, with the addition of a decision dialog on top.

	\vspace{-\baselineskip}
	\section{Survey}
	
	Each user who completed the experiment was presented with a set of multiple choice survey questions, in addition to a free form to provide feedback at the end. In the following, we discuss the most important results based on users' answers.
	Figure~\ref{fig:understanding} shows that although the majority of users understand what the text of the different Google permissions means, at least one-fourth of users expressed that they do not fully understand these permissions.
	Figure~\ref{fig:intuitive} allowed us to verify whether the experimental permission interfaces were intuitive to the users. More than 90\% of users answered affirmatively, indicating that our experiments' interface was user-friendly. Figure~\ref{fig:surprised} showed that the users in the \textit{FR} group were the ones that expressed the most surprise at what the apps can know about them, which is justified given the low Acceptance Likelihood in this group. Finally, more than 90\% of users (and 100\% of the \textit{FR} group) expressed interest in using a similar interface to the one they saw in the experiment (Figure~\ref{fig:realApps}). Overall, the survey results were in line with the experimental findings. Furthermore, surveyed users expressed the interest in \textit{``adding recommendations for whether one should install 3rd party apps''}, in \textit{``implementing similar functionalities in the Google Play Store and iOS App Store''}, and in \textit{``highlighting apps that actually misbehave rather than only the over-privileged ones''}. These ideas can potentially be realized in future works.

\end{document}